\documentclass[12pt]{article}

\usepackage{amssymb}
\usepackage{amsmath}
\usepackage{booktabs}
\usepackage{multirow}
\usepackage{makecell}
\usepackage{algorithm}
\usepackage[noend]{algpseudocode}
\usepackage{rotating}

\usepackage[colorlinks=true, linkcolor=red, citecolor=blue, filecolor=black, urlcolor=black]{hyperref}

\usepackage[a4paper, top=0.5in, bottom=0.5in, left=0.5in, right=0.5in]{geometry}
\usepackage{float}
\usepackage[font=small]{caption}
\usepackage{graphicx}

\title{Dynamic Object Geographic Coordinate Recognition: An Attitude-Free and Reference-Free Framework via Intrinsic Linear Algebraic Structures}

\author{
    Junfan Yi$^{1,2,\dag}$, 
    Ke-ke Shang$^{1,\dag,*}$\thanks{Corresponding author: \texttt{kekeshang@nju.edu.cn}}, 
    Michael Small$^{3}$\\
    $^{1}$Computational Communication Collaboratory, \\Nanjing University, Nanjing, 210023, P.R. China\\
    $^{2}$School of Geography and Ocean Science, \\Nanjing University, Nanjing, 210023, P.R. China\\
    $^{3}$Complex Systems Group, Department of Mathematics and Statistics, \\The University of Western Australia, Crawley, WA 6009, Australia
}

\date{Aug 2024}

\begin{document}
\maketitle

\noindent {\small 
\textsuperscript{\dag} These authors contributed equally to this work.}

\section{Abstract}

The Earth, a temporal complex system, is witnessing a shift in research on its coordinate system, moving away from conventional static positioning toward embracing dynamic modeling. Early positioning concentrates on static natural geographic features, with the emergence of geographic information systems introducing a growing demand for spatial data, the focus turns to capturing dynamic objects.
% which typically relies on expensive attitude measurement devices or external calibration objects.
However, previous methods typically rely on expensive devices or external calibration objects for attitude measurement.
We propose an applied mathematical model that utilizes time series, the nature of dynamic object, to determine relative attitudes without absolute attitude measurements, then employs Singular Value Decomposition (SVD)-based methods for 3D coordinate recognition.
The model is validated with negligible error in a numerical simulation, which is inherent in computer numerical approximations.
% We propose an applied mathematical model {\color{blue}这里用model吧，就说清楚数学，工程场景再提framework好了} that utilizes time series, the nature of dynamic object, to determine relative attitudes without absolute attitude measurements，
% {\color{blue}这句话也不对，越改越差，跟上一句话大部分都重了，你这相当于前四句，改掉了一句，后面两句也改差了}then employs Singular Value Decomposition (SVD)-based methods to {\color{red} 这里怎么写比较好，要确认一下,完成坐标识别？构建3D坐标？还是什么。}3D coordinate recognition .
% {\color{blue}Validation through 这句原话是什么？写的比这个好，不要用through。 a numerical simulation shows negligible error, which is inherent in computer numerical approximations.}
% To further validate our model, we propose a framework applied to an engineering scenario with only three cameras
%表达出我们想让这个模型落地实际场景的想法
%还少了使用UAV作为研究对象，不然怎么会出现检测，检测谁。
%第一句定位position改为capture
What in follows, to assess our model in the %{\color{red} the 更好，你是为了检验一个通用的工程场景，后面用了UAV做一个case,表达了一个的意思，不知道会不会让你测试鸟类，（救命）如果审稿人读来读去发现问题的话} 
engineering scenario,
%可以表达出我们使用UAV是因为UAV可以被我们操控，用于模拟任何运动的物体的意思 他们有养鸽子 然后给鸽子定位的 还发了蛮好的文章 麦克的学生养过 那个学生还在学校，现在是博士后的职位 鸽子更具有随机性 研究集群运动的 你这个就这样吧 后面你要是在那边也可以飞鸽子？好吧
%To {\color{blue}implement或adapt} our model {\color{blue}in或for} an engineering scenario
%{\color{red}好像也不行，是要写一句话在前面，但是这个句话写的不成功，implement或adapt都不太合适，要表达的意思是在具体的工程实践中测试我们数学方法的可行性}, 
we propose a framework featuring the integration of applied mathematics with artificial intelligence (AI), utilizing only three cameras to capture an unmanned aerial vehicle (UAV).
% 我觉得可以先不说利用人工智能，先说需要目标2d定位（可以） 是这样 但是你后面写了accurate 2D coordinate acquisitions，重复了 或者直接说 我们首先要对获取的图像进行预处理，这样避开重复讲一句话，也说明了是预处理 是上面一句已经说了相机 那么下面出现获取的图像是顺利成章的事情
% 尽量不要搞这种长难句which necessitates {\color{blue}the application of} artificial intelligence (AI) for camera-captured dynamic object detection as a preprocessing step. 你这里这么些太复杂了点，表达的很多，但这一块就不是重点，我们要表达出成体系了，但是AI不是优势，
We enhance the state-of-the-art You Only Look Once version 8 (YOLOv8) model by leveraging time series for the accurate 2D coordinate acquisitions, which is then used as input for 2D-to-3D conversion via our mathematics model. As a result, the framework achieves an Root Mean Square Error (RMSE) of \(4.9436 \, \text{m}\), an Mean Absolute Error (MAE) of \(4.7948 \, \text{m}\), a maximum error of \(8.8485 \, \text{m}\), and an R-squared of \(0.9567\), all of which showcase the high precision of our framework.
%{\color{blue}in final 3D coordinate recognition} .
% {\color{blue}The proposed framework is mathematically error-free, with all errors attributable to equipment inaccuracies and the limitations of AI, despite our AI-based detection being based on an enhanced version of the state-of-the-art YOLOv8 model.懂了，确实}
% It is crucial to emphasize that our mathematical method is error-free, with all errors stemming from external devices and our AI-based 2D coordinate acquisition, which is an advanced version of the current acknowledged best.
It is crucial to emphasize that our mathematical method is error-free, with any errors solely attributable to external devices or our AI-based 2D coordinate acquisition, which is an improved version of the currently acknowledged best method.
Our framework enriches geodetic theory by providing a streamlined model for the 3D positioning of non-cooperative targets, minimizing input attitude parameters, leveraging applied mathematics and AI.

\section{Introduction}

% 要记住几个关键词：地理，位置，计算机，应用数学，时间序列

% 先写定义，前两句话宏观，后面再细节accurate

% 第一段：同一个文献可以多引用几次
% 从位置的前沿引入（环境，气候变化，城市规划，公共健康）（自动驾驶，智慧城市，灾害管理，导航）。它的核心是Geodetic positioning，它是Geomatics的基本
% 起源于几何学的古老问题，Geodetic positioning 经过了从主要以二维测量为基础的传统地面测量方法，到现代三维空间测量的转变，成为测绘学（Geomatics）中不可或缺的重要组成部分，为资源管理、气候变化应对、智慧城市建设等提供了可靠的地理信息。
% {\color{red} 这三个地方的参考文献，需要把前面相关的也加上去，到时候如果参考文献超了，适当删减这里和前面的一般刊物的}

Geodetic positioning emerged from ancient geometric problems and subsequently evolved into a core technology of geomatics, applicable to any location worldwide, unrestrained by geographical boundaries. This evolution grants the technology systematic, dynamic, and interdisciplinary features \cite{chen2023iterative}, naturally leading to the leveraging of insights from physics \cite{hess2022physically,Balaian2024}, mathematics \cite{gomarasca2009basics,wayman1959least,yang2001adaptively}, with a growing emphasis on artificial intelligence \cite{gentine2018could,chen2023artificial,han2018solving,irrgang2021towards}. Crucially, rooted in both geometric challenges and spatial exploration demand, geodetic positioning is transforming from traditional ground-based methods to advanced spatial measurements. 
This transition, primarily from static geodetic mapping to dynamic 3D positioning, is driven by applied mathematics and computational science, enriches geospatial information, which is essential for global environment protection \cite{zhang2024water}, climate change mitigation \cite{gentine2018could,lin2012physically}, and city sustainability\cite{Balaian2024,ao2024national}, and further understand the complex earth system\cite{hess2022physically,gentine2018could,irrgang2021towards,reichstein2019deep}.

\begin{figure}[htbp] % h 表示在当前位置插入图片
    \centering % 图片居中
    \includegraphics[width=\linewidth]{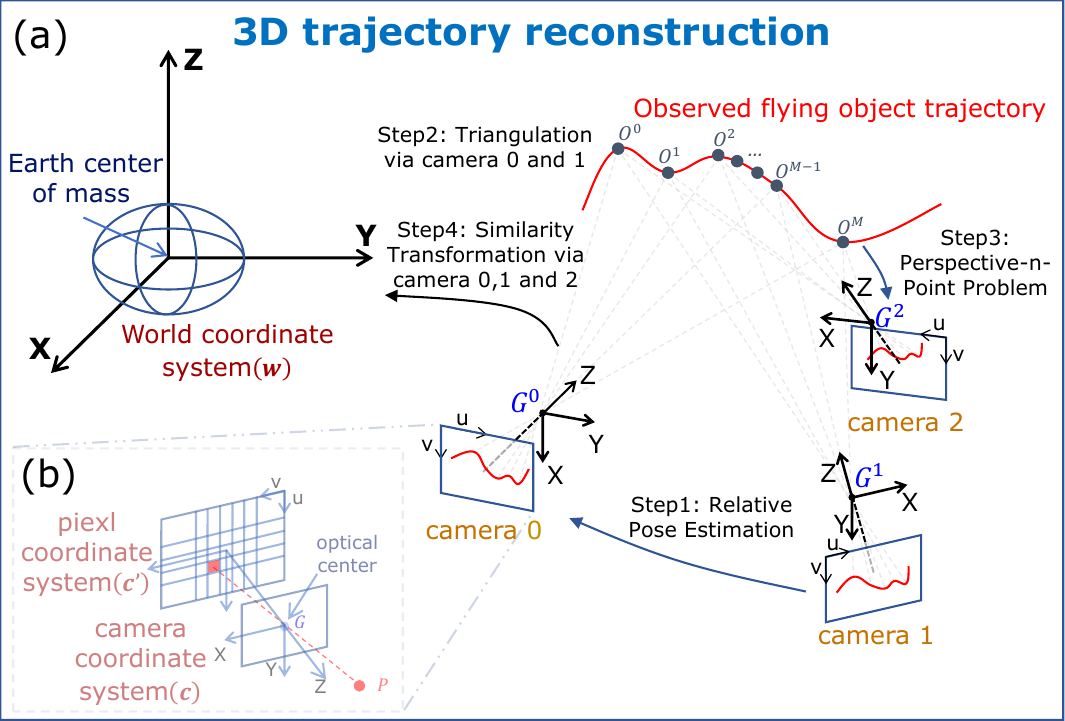} % 指定图片文件名，不需要文件扩展名
    \caption{
    In this paper, we utilize only three cameras to capture the trajectory of a flying object without measuring the attitudes of the cameras using calibration objects or devices. As shown in (a), 
    step 1: Three ground-based cameras are directed towards the flying object, capturing its trajectory denoted as \(O^j_{{c'}^i}\) at time step \( j \) in the pixel coordinate systems of camera \(i\) (\({c'}^i\)). The projection of a point \(P\) in 3D space onto the 2D pixel coordinate system is described as the pinhole camera model (b). Upon acquiring the 2D trajectories \(O^j_{{c'}^0}\) and \(O^j_{{c'}^1}\) in the pixel coordinate systems of camera 0 (\({c'}^0\)) and camera 1 (\({c'}^1\)), we employ a relative pose estimation method to calculate the relative pose of camera 1 with respect to camera 0.
    Step 2: With the 2D trajectories \(O^j_{{c'}^0}\) and \(O^j_{{c'}^1}\), and the determined pose of camera 1 relative to camera 0 from step 1, triangulation is used to calculate the 3D position of \(O^j\) in the camera 0 coordinate system (\({c}^0\)).
    Step 3: Obtained the trajectory of \(O^j\) in the camera 0 coordinate system (\({c}^0\)), we solve the Perspective-n-Point (PnP) problem to calculate the relative pose of camera 2 with respect to camera 0.
    Step 4: Given the coordinates of the three cameras \(G^i\) in both the world coordinate system (\(w\)) and the camera 0 coordinate system (\({c}^0\)), we solve for the similarity transformation between these two coordinate systems to compute the trajectory of \(O^j\) in the world coordinate system (\(w\)).
    }
    \label{fig:multiple-coordinate-system-transformation}
\end{figure}

Currently, 3D positioning systems can be classified into two categories, active and passive.
Active techniques, such as Global Navigation Satellite System (GNSS) \cite{hofmann2007gnss}, Inertial Navigation System (INS) \cite{barshan1995inertial} and Ultra Wide Band (UWB) \cite{yang2004ultra},
involve the target actively determining its own 3D position by carrying signal reception devices.
In contrast, passive methods, such as Light Detection and Ranging (LiDAR) \cite{wehr1999airborne}, photogrammetry \cite{mikhail2001introduction}, and Synthetic Aperture Radar (SAR) \cite{moreira2013tutorial}, are used to determine the 3D position of non-cooperative targets, typically relying on external calibration using calibration objects or expensive attitude measurement devices.
However, these approaches face significant challenges in scenarios involving non-cooperative targets, and in the absence of ground calibration objects and attitude measurement devices.
In dynamic real-world environments, the trajectory of the spatial target is captured as a position-time series across multiple sensors.
Matching these sequential observations, geometric transformations can compute the relative poses (attitudes and positions) of cameras.
In this perspective, we propose an efficient and simplified camera-based optical measurement rooted in applied mathematics which utilizes captured 2D coordinate time series to establish relative relationships between multiple cameras, and employ Singular Value Decomposition (SVD) to derive {\color{red} 3D} coordinate recognition (\autoref{fig:multiple-coordinate-system-transformation}; Section~\ref{sec:computation principle for world coordinate calculation}).

We conduct a numerical simulation in a \(200 \times 200 \times 100 \, \text{m}\) virtual 3D space.
Here, three ground-based cameras with unknown attitudes capture a flying object, represented as a 2D point in the images.
The captured 2D coordinate time series is then used to estimate the relative poses between the three cameras using SVD (Section~\ref{sec:computation principle for world coordinate calculation}). 
We combine the estimated poses with the 3D coordinates of the cameras to solve for the similarity transformation, which is subsequently used for the 3D coordinate recognition of the flying object.
The resulting errors are negligible, primarily due to inherent numerical approximations in the simulation.
Additionally, we consider engineering scenarios by introducing camera positioning errors, 2D coordinate deviations, and scene scaling factors to evaluate their impact on 3D coordinate recognition (\autoref{fig:reconstruction-error}). 
We find that 2D coordinate deviations have a greater impact on the resulting errors compared to camera positioning errors. 
Moreover, a larger scene size amplifies the influence of both camera positioning errors and 2D coordinate deviations on 3D coordinate recognition.
Nevertheless, the accuracy of 3D coordinate recognition remains within acceptable limits for general camera positioning and pixel deviation levels (Section~\ref{sec:numerical-simulations}).
Based on the 2D coordinate time series captured by multiple cameras, our mathematical approach for 3D coordinate recognition is precise, straightforward, and theoretically robust.
This method efficiently converts 2D coordinates to 3D coordinates, making attitude measurement unnecessary, with any errors arising solely from numerical approximations in simulations. 
Rather than relying on artificial intelligence, which often involves complex model architectures, substantial computational resources, and meticulous parameter tuning, this approach leverages explainable mathematical principles.

Furthermore, to validate the proposed method in a real-world scenario, we conduct a UAV experiment (Section~\ref{sec:UAV-case}) within a \(100 \times 100 \times 30 \, \text{m}\) area, using a UAV as the flying object and employing three cameras for 3D coordinate recognition. 
In engineering applications, it is essential to establish a system that links object detection to real-world 3D coordinate recognition. 
Leveraging advancements in artificial intelligence \cite{chen2023iterative}, we efficiently detect the UAV as a bounding box and determine its centroid to derive 2D coordinates, which are then converted into 3D coordinates using our mathematical approach.
Here, we employ YOLOv8 (You Only Look Once, version 8; Section~\ref{sec:UAV-detection}; Supplementary Information Section~\ref{sec:YOLOv8-structure}), a top-performing object detection method based on Convolutional Neural Networks (CNNs), to extract the detection bounding box of the UAV from the image. 
However, YOLOv8 still struggles with missed and false detections of small objects in complex scenarios \cite{liu2021survey}. 
To address these challenges, we propose YOLOv8-Time Series (YOLOv8-TS; Section~\ref{sec:yolov8-TS}), which leverages the temporal sequence and the physical characteristics of UAV motion, particularly its velocity, to effectively reduce missed and false detection.
Using the detected 2D coordinate time series from the three cameras, we follow the steps outlined in simulation (Section~\ref{sec:numerical-simulations}) for 3D coordinate recognition. 
Finally, the results of the 3D coordinate recognition show an RMSE of \(4.9436 \, \text{m}\), an Mean Absolute Error (MAE) of \(4.7948 \, \text{m}\), a maximum error of \(8.8485 \, \text{m}\), and an R-squared value of \(0.9567\) in a \(100 \times 100 \times 30 \, \text{m}\) scenario, demonstrating the robustness and effectiveness of our proposed time series-based method without prior knowledge of camera attitudes.

In all, we establish a 3D coordinate recognition system for non-cooperative targets without attitude measurement that is centered on applied mathematics, supplemented by AI-driven 2D detection technology.
The core of our system, which involves the transformation from 2D coordinates to 3D coordinates, leverages two key theories, applied time series and singular value decomposition, which ensure an error-free transition from 2D to 3D.

 \begin{figure}[htbp] % h 表示在当前位置插入图片
    \centering % 图片居中
    \includegraphics[width=\linewidth]{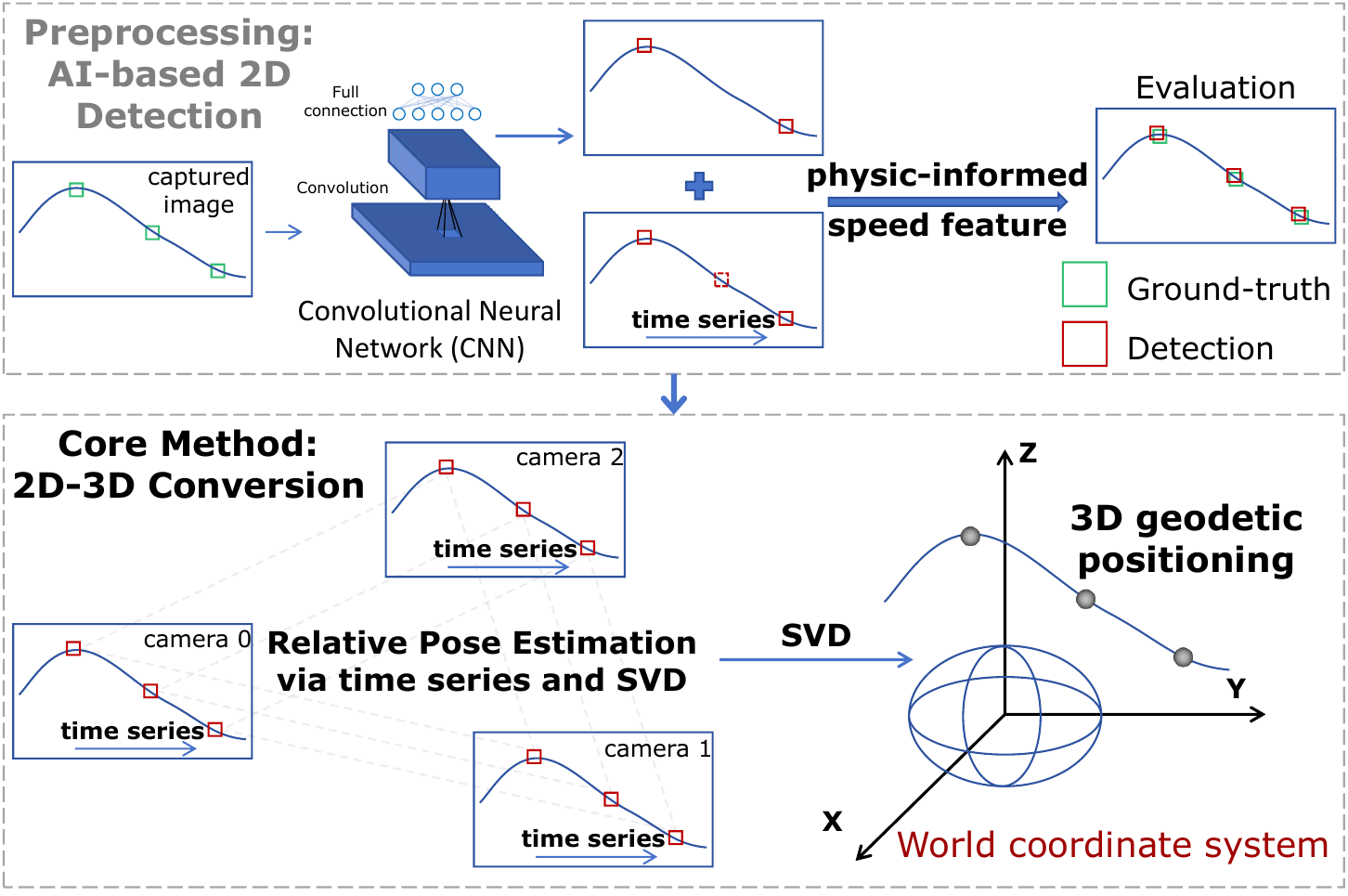} % 指定图片文件名，不需要文件扩展名
    \caption{
    % {\color{red} Auxiliary Step 改成Preprocessing吧，Auxiliary Step 是汽车的辅助踏板的意思，应该会有歧义。再就是 Preprocessing: AI-based 2D Detection改为灰色字体。区分一下重要程度 Relative Pose Estimation with time series 改成 Relative Pose Estimation via time series and SVD
    % 基于时间序列与应用数学的原理的2d-3d conversion。
    % 利用ai辅助，得到目标2d，基于时间序列结合物理特征：速度，对结果改进main-principle}    
    The 2D-to-3D conversion framework. 
    The framework begins with AI-based 2D detection using a Convolutional Neural Network (CNN) to obtain the 2D coordinate time series of the object in the captured images. 
    These coordinates are then refined using time series and the physical feature of velocity.
    The core method utilizes time series and Singular Value Decomposition (SVD) to estimate the relative poses of the cameras. 
    An SVD-based approach is further employed to calculate the similarity transformation matrix, deriving the camera-to-world coordinate transformation and ultimately achieving 3D geodetic positioning in the world coordinate system.
    }
    \label{fig:main-principle}
\end{figure}

\section{Numerical simulations of coordinate recognition
\label{sec:numerical-simulations}
}
% 承接上文说的数学方法，接下来的仿真。不停强调我的优势。紧密围绕2d到3d的数学原理。

As depicted in the mathematical model of Section~\ref{sec:computation principle for world coordinate calculation}, 
the unique time series of 2D coordinates formed by the trajectory of a moving spatial point, captured by cameras with known 3D world coordinates, provides relative position information of points. When this information is processed using Singular Value Decomposition (SVD), which extracts singular vectors from non-square matrices, it helps identify the relative poses of the cameras. Following this process, once the 3D world coordinates and the relative poses of the cameras are obtained, SVD can further facilitate both 2D to 3D and 3D to 3D coordinate transformations of observed points. Notably, this method does not require prior knowledge of camera {\color{red} attitudes} for reducing the dependence on external devices.

\begin{figure}[htbp]
  \centering
  \includegraphics[width=0.8\textwidth]{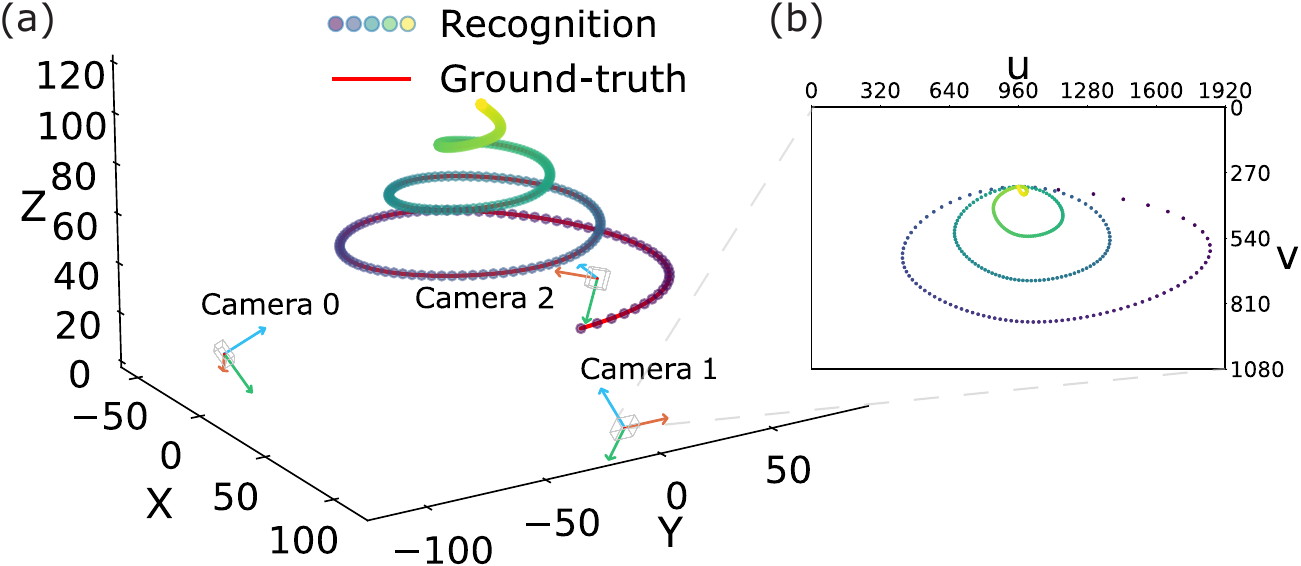}
  \caption{The 3D scene of the numerical simulation experiment. 
  (a) The red ascending spiral line from \((70, 0, 30)\,\text{m}\) to \((0,0,100)\,\text{m}\) represents the trajectory formed by the ground-truth spatial points over 300 time steps. The gradient-colored dots, based on the viridis colormap, represent the calculated spatial points throughout the same period. The colors transition from dark purple to shades of blue and green, finishing with bright yellow, indicating the progression over time.
  Three ground cameras, each located 100 m from the center \((0,0,0)\,\text{m}\), are positioned at 120° intervals. These cameras, represented by 3D coordinate axes, track the moving point and capture its coordinates in their respective 2D pixel coordinate systems. 
  The optical centers of three cameras in the world coordinate system are \((\text{-}50.0, \text{-}86.6, 0.0)\,\text{m}\), \((100.0, 0.0, 0.0)\,\text{m}\), and \((\text{-}50.0, 86.6, 0.0)\,\text{m}\) respectively. Each camera coordinate system is aligned with the X-axis (red) parallel to the ground and the Z-axis (blue) pointing to \((0, 0, 65)\,\text{m}\), forming a right-handed system with the Y-axis (green). The cameras are distortion-free, with focal lengths \(f_x = f_y = 1000 \, \text{pixels}\), principal points at \(c_x=960 \, \text{pixels}\), \(c_y=540 \, \text{pixels}\), and pixel skew \(s=0\). (b) The projection of the spatial point onto the image plane of camera, with axes \(u\) and \(v\) representing the horizontal and vertical pixel axes respectively.}
  \label{fig:SimulationsScene}
\end{figure}

We conduct a numerical simulation using three non-aligned ground cameras and a moving point over 300 time steps within a virtual 3D space measuring \(200 \times 200 \times 100\) m (\autoref{fig:SimulationsScene}(a)). 
The cameras are used to capture the 2D coordinates of the moving point at each time step. Additionally, the 3D coordinates of each camera in the world coordinate system are known.

First, leveraging the time series of 2D coordinates from camera 0 and camera 1 over 300 time steps (with 8 time steps being the theoretical minimum, see Section~\ref{sec:computation principle for world coordinate calculation}), we obtain 300 pairs of 2D coordinates of the moving spatial point.
These pairs are used with SVD to extract the relative poses of the camera 0 and camera 1 (\autoref{fig:multiple-coordinate-system-transformation}(a); Section~\ref{sec:computation principle for world coordinate calculation}, \autoref{eq:essential-matrix}). 
Second, with the calculated relative poses of these two cameras, we construct rays from the optical center to the 2D pixel points in both cameras. Applying SVD to minimize the distance between the target 3D point and the two rays, we obtain the 3D coordinates of the spatial point in the coordinate system of camera 0 (\autoref{fig:multiple-coordinate-system-transformation}(b); Section~\ref{sec:computation principle for world coordinate calculation}, \autoref{eq:triangulation-O}).
Third, for camera 2, based on the calculated 3D coordinates of the spatial point in the camera 0 coordinate system and its 2D coordinates in the pixel coordinate system of camera 2, we use the Effective Perspective-n-Point (EPnP) (Section~\ref{sec:computation principle for world coordinate calculation}) to calculate the relative pose of camera 2 with respect to camera 0 (\autoref{fig:multiple-coordinate-system-transformation}(c)).
Based on the calculated poses of the three cameras relative to camera 0, we determine the coordinates of each camera in the camera 0 coordinate system (Section~\ref{sec:computation principle for world coordinate calculation}, \autoref{eq:camera-coordinate-in-camera0}).
% {\color{green}
% Finally, using the calculated poses of the three cameras relative to camera 0, we determine the coordinates of each camera in the camera 0 coordinate system (Section~\ref{sec:computation principle for world coordinate calculation}, \autoref{eq:camera-coordinate-in-camera0}).利用计算得到的三台相机在相机0坐标系下的坐标和已知的在世界坐标系下的坐标 ，we then apply Singular Value Decomposition (SVD) to derive the similarity transformation matrix between these coordinate systems. With this matrix, we compute the spatial point coordinates in the world coordinate system over 300 time steps (see \autoref{fig:multiple-coordinate-system-transformation}(d); Section~\ref{sec:computation principle for world coordinate calculation}, \autoref{eq:camera-coordinate-in-camera0} to \autoref{eq:object-coordinate-world}).
% Finally,由计算得到的三台相机相对于相机0的姿态，每台相机在相机0坐标系下的坐标可以得到(Section~\ref{sec:computation principle for world coordinate calculation}, \autoref{eq:camera-coordinate-in-camera0})，这样我们同时得到了三台相机在相机0坐标系和世界坐标系下的坐标，它们被利用SVD to obtain the similarity transformation matrix between 相机0 coordinate systems and 世界 coordinate systems。这样，我们就可以利用这个相似变换矩阵得到the coordinates of the spatial point in the world coordinate system for 300 time steps ( \autoref{fig:multiple-coordinate-system-transformation}(d) ; Section~\ref{sec:computation principle for world coordinate calculation}, \autoref{eq:camera-coordinate-in-camera0} to \autoref{eq:object-coordinate-world} ).}
Finally, the 3D coordinate pairs of the spatial point in the coordinate system of camera 0 and the world coordinate system can be represented by three cameras in parallel, then we apply SVD to obtain the similarity transformation matrix between these two coordinate systems, thus obtaining the 3D coordinates of the moving spatial point in the world coordinate system ( \autoref{fig:multiple-coordinate-system-transformation}(d) ; Section~\ref{sec:computation principle for world coordinate calculation}, \autoref{eq:camera-coordinate-in-camera0} to \autoref{eq:object-coordinate-world} ).

The metrics we report include Root Mean Square Error (RMSE), Mean Absolute Error (MAE), Maximum Error, and R-squared (Section~\ref{sec:metrics}). The results of spatial point coordinate recognition (\autoref{fig:SimulationsScene}) show that errors are almost zero, which are considered negligible in a \(200 \times 200 \times 100\) m scene: the RMSE is \(3.2 \times 10^{\text{-}4}\) m, the MAE is \(3.2 \times 10^{\text{-}4}\) m, the Maximum Error is \(4.9 \times 10^{\text{-}4}\) m, and the R-squared is almost 1. 
This demonstrates that our theoretical method is virtually error-free, showcasing the value of applied mathematics in providing precise calculations, in contrast to methods that rely on artificial intelligence for spatial point coordinate estimation.
Actually, the primary source of pure simulation error is from the approximation inherent in computer numerical solutions.

\begin{figure}[htbp]
  \centering
  \includegraphics[width=\textwidth]{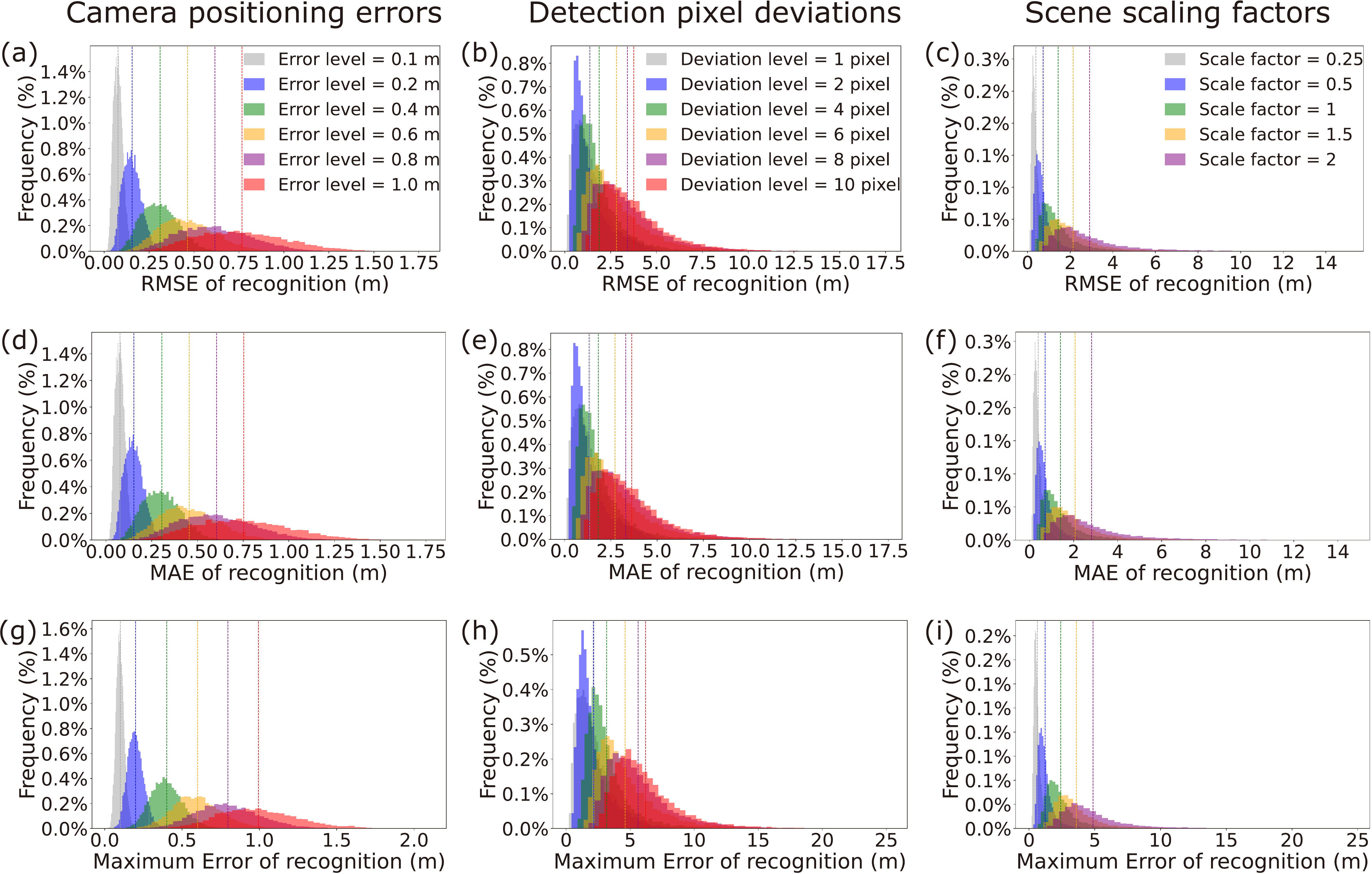}
  \caption{
  Error histograms of coordinate recognition under different perturbations. 
  For each perturbation level, 10,000 experiments are conducted, resulting in RMSE, MAE, and Maximum Error. These errors are displayed as histograms, where the horizontal axis represents the metric, divided into 50 equal-width intervals based on the maximum and minimum values. The vertical axis shows the frequency of errors within each interval. 
  Dashed lines indicate the mean errors from the 10,000 results at each perturbation level.
  (a), (d), (g) Camera positioning errors: Histograms show the distribution of coordinate recognition errors across 10,000 simulations for each positioning error level (\( e_p = 0.1, 0.3, 0.5, 0.7, 0.9, 1.1 \) m) for RMSE (a), MAE (d), and Maximum Error (g).
  (b), (e), (h) Pixel deviations: Histograms display the distribution of coordinate recognition errors across 10,000 simulations for each pixel deviation level (\( e_{\text{pix}} = 1, 2, 4, 6, 8, 10 \) pixels) for RMSE (b), MAE (e), and Maximum Error (h).
  (c), (f), (i) Scene scaling factors: The scene is scaled using factors of 0.25, 0.5, 1, 1.5, and 2, with a camera positioning error of 0.2 m and a random pixel deviation of 3 pixels, for RMSE (c), MAE (f), and Maximum Error (i). 
  A scaling factor of 0.25 implies that the coordinates of both the spatial points and the camera positions are scaled to 25\% of their original values. The cameras are oriented with the X-axis parallel to the ground and the Z-axis pointing towards the spatial coordinate where \(x = 0\), \(y = 0\), and the Z-value is the average of the maximum and minimum Z-coordinates of the spatial point over 300 time steps. For each simulation, the ground-truth 3D coordinates of the moving spatial point and the camera positions are adjusted according to the scaling factor.
  }
  \label{fig:reconstruction-error}
\end{figure}

However, in engineering scenarios, various factors such as camera positioning errors, pixel detection deviations and scaling in experimental scene size will impact the accuracy of 3D coordinates recognition. As shown in \autoref{fig:reconstruction-error}, to evaluate those factors we perform further simulations.

\textbf{Camera positioning errors.}
We firstly explore the impact of camera positioning errors on the accuracy of 3D spatial point coordinate recognition by introducing random perturbations to each camera location. In particular, we consider six levels of positioning errors: \(e_p = 0.1, 0.3, 0.5, 0.7, 0.9, 1.1\,\text{m}\)  (Supplementary Information Section~\ref{sec:Parameter-selection-of-simulation-experiment}). 
For each level, the camera positions are perturbed by adding a randomly generated error vector to original positions:

\begin{equation}
G_{w,\text{perturbed}}^i = G_{w}^i + \mathbf{E}_p^i,
\end{equation}

where \(G_{w}^i\) denotes the original position of camera \(i\) in the world coordinate system, \(G_{w,\text{perturbed}}^i\) represents the perturbed position, and \(\mathbf{E}_p^i\) is a \(3 \times 1\) vector 
in which each element is a random positioning error uniformly distributed within \([\text{-}e_p, e_p]\).

\autoref{fig:reconstruction-error}(a) illustrates the result of the calculated coordinates for each level of camera positioning error. 
For 3D coordinate recognition with camera positioning deviations of \(0.1 \, \text{m}\), \(0.3 \, \text{m}\), \(0.5 \, \text{m}\), \(0.7 \, \text{m}\), \(0.9 \, \text{m}\), and \(1.1 \, \text{m}\), the RMSE are \(0.0777 \, \text{m}\), \(0.1553 \, \text{m}\), \(0.3109 \, \text{m}\), \(0.4642 \, \text{m}\), \(0.6162 \, \text{m}\), and \(0.7673 \, \text{m}\), respectively, while the MAE are \(0.0763 \, \text{m}\), \(0.1525 \, \text{m}\), \(0.3053 \, \text{m}\), \(0.4558 \, \text{m}\), \(0.6050 \, \text{m}\), and \(0.7533 \, \text{m}\), respectively. The Maximum Error for these deviations are \(0.1004 \, \text{m}\), \(0.2004 \, \text{m}\), \(0.4012 \, \text{m}\), \(0.6004 \, \text{m}\), \(0.7965 \, \text{m}\), and \(0.9936 \, \text{m}\), respectively. 
As the camera positioning error increases, the error in spatial point coordinate recognition also increases.

% Furthermore, \(95\%\) of the calculated coordinates have RMSEs within \(0.0707 \, \text{m}\), \(0.1409 \, \text{m}\), \(0.2804 \, \text{m}\), \(0.4257 \, \text{m}\), \(0.5601 \, \text{m}\), and \(0.6961 \, \text{m}\) for each level, respectively.
% {\color{blue}As the camera positioning error increases, the error in spatial point coordinate recognition also increases.}

% rmse：0.0777 0.1553 0.3109 0.4642 0.6162 0.7673 
% mae：0.0763 0.1525 0.3053 0.4558 0.6050 0.7533 
% maxerror：0.1004 0.2004 0.4012 0.6004 0.7965 0.9936 
% 95：0.1224 0.2441 0.4857 0.7373 0.9702 1.2057

\textbf{Detection pixel deviations.}
We then investigate the effect of 2D pixel coordinate deviations on the accuracy of {\color{red}3D} coordinate recognition. Let \(e_{\text{pix}}\) represent the number of pixel deviation. We consider six levels of pixel deviation: \(e_{\text{pix}} = 1, 2, 4, 6, 8, 10\,\text{pixels}\) (Supplementary Information Section~\ref{sec:Parameter-selection-of-simulation-experiment}).
In each simulation, the 2D pixel coordinates of the spatial point in each image from the three cameras are perturbed by adding a randomly generated error vector to the observed point at each timestamp:

\begin{equation}
O_{c'^i,\text{perturbed}}^j = O_{c'^i}^j + \mathbf{E}_{\text{pix}}^j,
\end{equation}

where \(O_{c'^i}^j\) denotes the original 2D coordinates of the observed spatial point \(O^j\) at time step \(j\) in the pixel coordinate system \(c'^i\), \(O_{c'^i,\text{perturbed}}^j\) represents the perturbed 2D coordinates, and \(\mathbf{E}_{\text{pix}}^j\) is a \(2 \times 1\) vector representing the random pixel deviation uniformly distributed within \([\text{-}e_{\text{pix}}, e_{\text{pix}}]\).

\autoref{fig:reconstruction-error}(b) shows the result of the calculated 3D coordinates of the spatial point for each level of pixel deviation. The RMSE for pixel deviations of \(1\), \(2\), \(4\), \(6\), \(8\), and \(10\) pixels are \(1.3811 \, \text{m}\), \(1.3465 \, \text{m}\), \(1.8805 \, \text{m}\), \(2.8155 \, \text{m}\), \(3.4120 \, \text{m}\), and \(3.7553 \, \text{m}\), respectively, while the MAE are \(1.3545 \, \text{m}\), \(1.3115 \, \text{m}\), \(1.8195 \, \text{m}\), \(2.7303 \, \text{m}\), \(3.3034 \, \text{m}\), and \(3.6272 \, \text{m}\), respectively. The Maximum Error for these pixel deviations are \(2.0960 \, \text{m}\), \(2.1552 \, \text{m}\), \(3.1625 \, \text{m}\), \(4.6152 \, \text{m}\), \(5.6120 \, \text{m}\), and \(6.2098 \, \text{m}\), respectively.
It is evident that pixel deviations have a much greater impact on spatial point coordinate recognition compared to camera positioning errors.

% Additionally, \(95\%\) of the calculated coordinates have RMSEs within \(1.8551 \, \text{m}\), \(2.0429 \, \text{m}\), \(2.3978 \, \text{m}\), \(3.4970 \, \text{m}\), \(4.0850 \, \text{m}\), and \(4.2947 \, \text{m}\), respectively.

% rmse:1.3811 1.3465 1.8805 2.8155 3.4120 3.7553 mae:1.3545 1.3115 1.8195 2.7303 3.3034 3.6272 max:2.0960 2.1552 3.1625 4.6152 5.6120 6.2098 
% r:0.9988 0.9987 0.9977 0.9951 0.9930 0.9917 95:3.2131 3.5384 4.1532 6.0570 7.0754 7.4387

\textbf{Scene scaling.}
We finally conduct a comprehensive study to assess the impact of scene scaling on the accuracy of 3D coordinate recognition, considering both camera positioning errors and pixel deviations. The scene is scaled using factors of 0.25, 0.5, 1, 1.5, and 2 (Supplementary Information Section~\ref{sec:Parameter-selection-of-simulation-experiment}). 
For each scaling factor, we scale the experimental scene (details of the scaling process are provided in \autoref{fig:reconstruction-error}(c)) while introducing a camera positioning error of 0.2 m and a random pixel deviation of 3 pixels.

\autoref{fig:reconstruction-error}(c) depicts the errors in the calculated 3D coordinates for each scaling factor, with RMSE of \(0.3942 \, \text{m}\), \(0.7316 \, \text{m}\), \(1.4404 \, \text{m}\), \(2.1382 \, \text{m}\), and \(2.9114 \, \text{m}\), and corresponding MAE of \(0.3829 \, \text{m}\), \(0.7088 \, \text{m}\), \(1.3940 \, \text{m}\), \(2.0694 \, \text{m}\), and \(2.8185 \, \text{m}\) for scaling factors of \(0.25\), \(0.5\), \(1\), \(1.5\), and \(2\), respectively. The Maximum Error for these scaling factors are \(0.6415 \, \text{m}\), \(1.2189 \, \text{m}\), \(2.4168 \, \text{m}\), \(3.5901 \, \text{m}\), and \(4.8698 \, \text{m}\), respectively.
When both camera positioning errors and pixel deviations are considered, the larger the spatial scene, the greater the coordinate recognition errors.
%{\color{blue}In summary, camera positioning errors have the smallest impact on spatial point coordinate recognition, while pixel deviations have the largest impact. The size of the spatial scene amplifies these errors further.}

% Furthermore, \(95\%\) of the calculated coordinates have RMSEs within \(0.4759 \, \text{m}\), \(0.9363 \, \text{m}\), \(1.8858 \, \text{m}\), \(2.8065 \, \text{m}\), and \(3.8478 \, \text{m}\), respectively.

% rmse:0.3942 0.7316 1.4404 2.1382 2.9114 mae:0.3829 0.7088 1.3940 2.0694 2.8185 max:0.6415 1.2189 2.4168 3.5901 4.8698 r:0.9999 0.9997 0.9986 0.9970 0.9944 
% 95:0.8242 1.6217 3.2663 4.8610 6.6647

All in all, our simulations show that using applied mathematics to locate spatial points results in virtually no error. All the errors we observe stem from equipment inaccuracies, the higher the precision of the equipment, the more accurate the spatial point coordinate recognition. Even with equipment errors, precision in applied math enables overcoming the challenges posed by defects.

\section{Coordinate recognition experiment: UAV case
\label{sec:UAV-case}
}
The UAV experiment is conducted at the Second Stadium of Xianlin Campus, Nanjing University (118.948°E, 32.123°N). The experimental setup includes a UAV, three cameras, and a GNSS receiver. The process for UAV coordinate recognition in the world coordinate system is depicted in \autoref{fig:UAV-experiment-schematic-diagram}.
Before implementing our mathematical model for 3D coordinate recognition, the first challenge is how to accurately obtain the 2D coordinates of the spatial point, which is essentially a target detection problem in computer vision. Thus, the experiment is structured into two parts: UAV detection, serving as a preliminary step to acquire the 2D coordinates vital for subsequent 2D-to-3D conversion; and UAV {\color{red}3D} coordinate recognition, which is the core of our study, relying on the application of mathematical principles for converting these 2D coordinates into 3D spatial points.
%Accurate detection is crucial because it forms the basis for applying the time-series analysis principles discussed in this paper to real-world coordinate recognition. 
It is important to note that, we consistently utilize the principles of time series to improve the accuracy of YOLOv8, the widely recognized top-performing UAV detection method based on Convolutional Neural Networks (CNNs), renaming it YOLOv8-TS (Time Series).

\subsection{UAV detection in time series images
\label{sec:UAV-detection}
}
\begin{figure}
    \centering
    \includegraphics[width=1\linewidth]{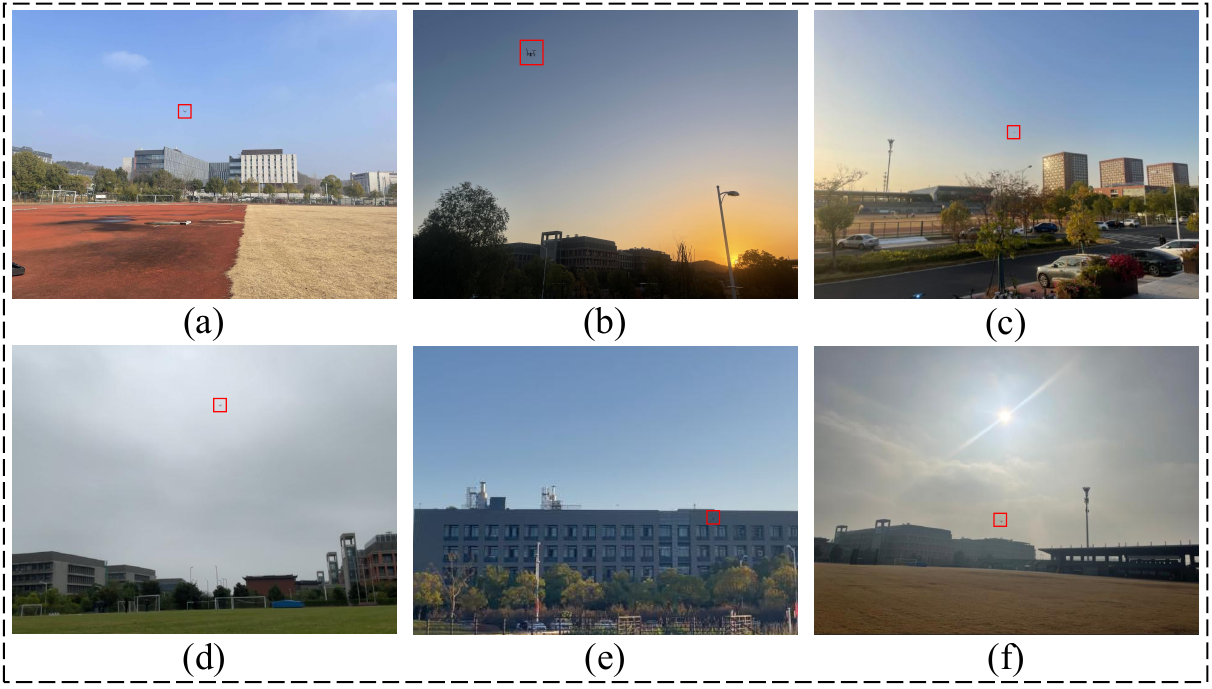}
    \caption{
    Images of a flying UAV captured by ground cameras under various background conditions, with UAV detections highlighted by red rectangles. Backgrounds include (a) clear, cloudless sky, (b) dusky, dark sky, (c) afternoon sky, (d) cloudy sky, (e) buildings in the background, and (f) backlighting. 
    % The UAV experiment setup, including a UAV, cameras, and a GNSS receiver used to measure the ground-truth positions of the cameras in the world coordinate system.
    }
    \label{fig:UAV-in-multiple-scenarios}
\end{figure}

\textbf{Data preprocessing.}
We record time-series images of a flying UAV for 300 seconds using three ground cameras. Each camera operates at a frame rate of 30 frames per second (fps), capturing a total of 9,000 images (fps $\times$ time). The three cameras capture 5,022, 5,706, and 6,157 images containing the UAV, respectively, as the UAV occasionally flies out of the cameras' field of view. To enhance UAV detection capabilities in complex sky backgrounds, additional images are captured in scenarios with backgrounds of buildings, twilight, and backlighting totally 487 images are added to UAV dataset (\autoref{fig:UAV-in-multiple-scenarios}). In total, we collect 17,370 images containing the UAV for model training. For the test set, we use 1,447 images from each of the three cameras, totaling 4,341 images. The ratio of the training set to the test set is 8:2 (\autoref{fig:UAV-experiment-schematic-diagram}(a)).

\textbf{
Model training.
}
For UAV detection in time series images, the YOLOv8 
(You Only Look Once, version 8; Section~\ref{sec:UAV-detection}; Supplementary Information Section~\ref{sec:YOLOv8-structure})
model is utilized as the detection framework (\autoref{fig:UAV-experiment-schematic-diagram}(b)). YOLOv8 is widely regarded as one of the best in the object detection field for its exceptional speed and accuracy. The model processes the entire image using a single neural network, simultaneously predicting bounding boxes and class probabilities. 

Among the five versions of YOLOv8—n, s, m, l, and x—YOLOv8n is selected in this study due to its lower computational cost and faster inference speed. YOLOv8n resamples the input image to a resolution of 640×640×3 and employs the Cross Stage Partial Darknet (CSPDarknet) backbone to generate a 20×20 feature map. 
In the neck, the model uses the Path Aggregation Feature Pyramid Network (PAFPN) structure to output features at scales of 80×80, 40×40, and 20×20. 
The head adopts a decoupled structure that independently optimizes classification and localization tasks, enhancing detection performance (Detailed information on the YOLOv8 model is provided in the Supplementary Information Section~\ref{sec:YOLOv8-structure}). 
We use the Stochastic Gradient Descent (SGD) algorithm for end-to-end network optimization, utilizing the official YOLOv8n as our pretrained weights (see Supplementary Information Section~\ref{sec:YOLOv8-structure} for more detailed training parameters).

\textbf{Prediction based on basic detection with YOLOv8.}
The YOLOv8 model is utilized to detect UAV in time-series images captured by three cameras, providing basic detection results with bounding boxes for each image at every time step (\autoref{fig:UAV-experiment-schematic-diagram}(c)). 
In the UAV experiment, the time-series images from these cameras, covering a total of 5,000 time steps, are specifically used for UAV coordinate recognition.

For each image, the model's predictions produce multiple bounding boxes with varying confidence levels. 
To filter potential UAV detections, a confidence threshold of 0.25 and an Intersection Over Union (IOU) threshold of 0.7 are applied, following the specifications in the YOLOv8 documentation. The confidence threshold sets the minimum confidence level required to accept detections, discarding those below this threshold. The IOU threshold is used for Non-Maximum Suppression (NMS) to reduce duplicate detections.
In this single-UAV scenario, only the detection with the highest confidence in each image is considered, even if multiple bounding boxes meet the thresholds. Across 5,000 images, cameras 1, 3, and 5 detect the UAV in 3,014, 1,975, and 2,352 images, respectively.

% Data collected from three cameras were used to predict the presence of UAVs. Each camera performed UAV detection on a continuous sequence of 5,000 frames within the same time period.

% The reconstruction of UAV coordinates was conducted using data collected from three cameras: camera 1, camera 3 and camera 5. The data used as input for the UAV detection model included:

% \begin{itemize}
%     \item camera 1: Frames 2000 to 7000, resulting in 3014 frames with UAV detection after YOLOv8 prediction, and 3604 frames after enhancement.
%     \item camera 3: Frames 5000 to 10000, yielding 1975 frames with UAV detection after YOLOv8 prediction, and 2584 frames after enhancement.
%     \item camera 5: Frames 7000 to 12000, providing 2352 frames showing UAV detection after YOLOv8 prediction, and 3134 frames after enhancement.
% \end{itemize}

\textbf{YOLOv8-TS and prediction improvement.}
When using the original YOLOv8 model for UAV detection, challenges such as false detections caused by interference from other moving objects, like birds and dragonflies, and missed detections in complex backgrounds are encountered. 
To mitigate these issues, we develop YOLOv8-Time Series (YOLOv8-TS) (\autoref{fig:UAV-experiment-schematic-diagram}(d); Section~\ref{sec:yolov8-TS}), 
a refined version of YOLOv8 that incorporates time-series analysis and physics-informed speed modeling.

This approach leverages the distinct physical characteristics of the UAV, particularly its speed, which differs from other flying objects within the field of view, as well as the consistent motion state of the UAV over short time sequences. YOLOv8-TS ultimately provides a result confirmed as a UAV, selecting from one of the three categories it assesses: confirmed as a UAV, confirmed as not a UAV, and pending confirmation.

\textbf{Model evaluation.}
We use \(IoU-P\), \(IoU-R\), \(IoU-F1\), and Area Under Curve \((AUC)\) as our evaluation metrics (as detailed in Section~\ref{sec:metrics}) to assess the detection performance of YOLOv8 and YOLOv8-TS, with the results presented in \autoref{tab:detection_model_result}.

\begin{table}[h]
    \centering
    \caption{
    Metrics for the YOLOv8 and YOLOv8-TS models. 
    YOLOv8 is widely recognized as one of the best object detection algorithms, known for its exceptional balance of efficiency and accuracy in image-based tasks, making it an ideal benchmark for evaluating the performance of the proposed YOLOv8-TS.
    }
    \begin{tabular}{ccccccccc}
        \hline
        Method & \(IoU-P\) & \(IoU-R\) & \(IoU-F1\) & \(AUC\)\\
        \hline
        YOLOv8 & \textbf{0.98}  & 0.82 & 0.89 & 0.88\\
        YOLOv8-TS & 0.97 & \textbf{0.94} & \textbf{0.95} & \textbf{0.97} \\
        \hline
    \end{tabular}
    \label{tab:detection_model_result}
\end{table}

Overall, YOLOv8-TS shows improvements over YOLOv8, particularly in \(IoU-R\), \(IoU-F1\), and \(AUC\). The \(IoU-P\) for YOLOv8-TS is 0.97, slightly lower than YOLOv8 (0.98). This decrease is attributed to the reduction in missed detections, which introduces a substantial number of correct UAV detections (true positives, TP) but also some incorrect UAV detections (false positives, FP), resulting in a slight drop in \(IoU-P\).
By reducing missed detections, previously undetected UAVs (false negatives, FN) are converted into TP or FP, leading to a significant increase in \(IoU-R\) from 0.82 in YOLOv8 to 0.94 in YOLOv8-TS, reflecting the improved ability to capture UAVs in various scenarios. The \(IoU-F1\), the harmonic mean of \(IoU-P\) and \(IoU-R\), increases from 0.89 in YOLOv8 to 0.95 in YOLOv8-TS, primarily due to the improvement in \(IoU-R\).
For AUC, YOLOv8-TS achieves 0.9746, nearly 0.1 higher than the 0.8793 of YOLOv8, indicating a significantly enhanced ability to distinguish UAV from their surroundings across various scenarios.

\subsection{UAV coordinate recognition}
\textbf{Coordinate recognition.} After preprocessing for UAV detection results (as detailed in Supplementary Information Section~\ref{sec:preparation-for-UAV-coordinate-recognition}), we obtain synchronized and aligned UAV 2D coordinate time series from all three cameras under a unified time framework for UAV coordinate recognition (\autoref{fig:UAV-experiment-schematic-diagram}(e); following the steps in Section~\ref{sec:numerical-simulations} of numerical simulations). We first use the UAV 2D coordinate time series from camera 0 and camera 1 to calculate the relative poses between these two cameras. Specifically, we identify the common time period between these two time series and pair the UAV 2D coordinates at each time step. These coordinate pairs from all time steps are used to calculate the relative pose of camera 1 with respect to camera 0 (\autoref{fig:multiple-coordinate-system-transformation}(a), step 1).

Using the calculated relative poses of camera 0 and camera 1 along with the UAV 2D coordinate time series from these two cameras, we establish the geometric relationship and use SVD decomposition to obtain the UAV 3D coordinates in the camera 0 coordinate system for each time step (\autoref{fig:multiple-coordinate-system-transformation}(a), step 2).

For the remaining cameras \(i\) (where \(N > i \geq 2\), with \(i = 2\) and \(N = 3\) in the UAV experiment), we use the preprocessed UAV 2D coordinate time series from camera \(i\) (Supplementary Information Section~\ref{sec:preparation-for-UAV-coordinate-recognition}) and the calculated 3D UAV coordinates in the camera 0 coordinate system to compute the relative pose of camera \(i\) with respect to camera 0 (\autoref{fig:multiple-coordinate-system-transformation}(a), step 3). 
Additionally, the 2D coordinate time series from camera \(i\) provide supplementary observations that are also used for UAV coordinate recognition(\autoref{fig:multiple-coordinate-system-transformation}(a), step 2; Supplementary Information Section~\ref{sec:preparation-for-UAV-coordinate-recognition}).

Based on the calculated poses of each camera relative to camera 0, we determine the coordinates of each camera within the camera 0 coordinate system. 
Considering the \(N\) cameras as \(N\) spatial points, we use their coordinates in the camera 0 coordinate system, along with their known coordinates in the world coordinate system obtained from the GNSS receiver, to calculate the similarity transformation matrix between the camera 0 coordinate system and the world coordinate system which is used for transforming the UAV 3D coordinate time series from the camera 0 coordinate system to the world coordinate system (\autoref{fig:multiple-coordinate-system-transformation}(a), step 4).

\textbf{Evaluation.}
To evaluate the results of UAV coordinate recognition ({\autoref{fig:UAV-experiment-schematic-diagram}(f)}), ground-truth UAV 3D coordinates are provided by an onboard GNSS receiver, which updates six times per second with a positioning accuracy of up to \(1 \, \text{m}\).

\begin{figure*}[htbp]
\centering
\includegraphics[width=0.5\linewidth]{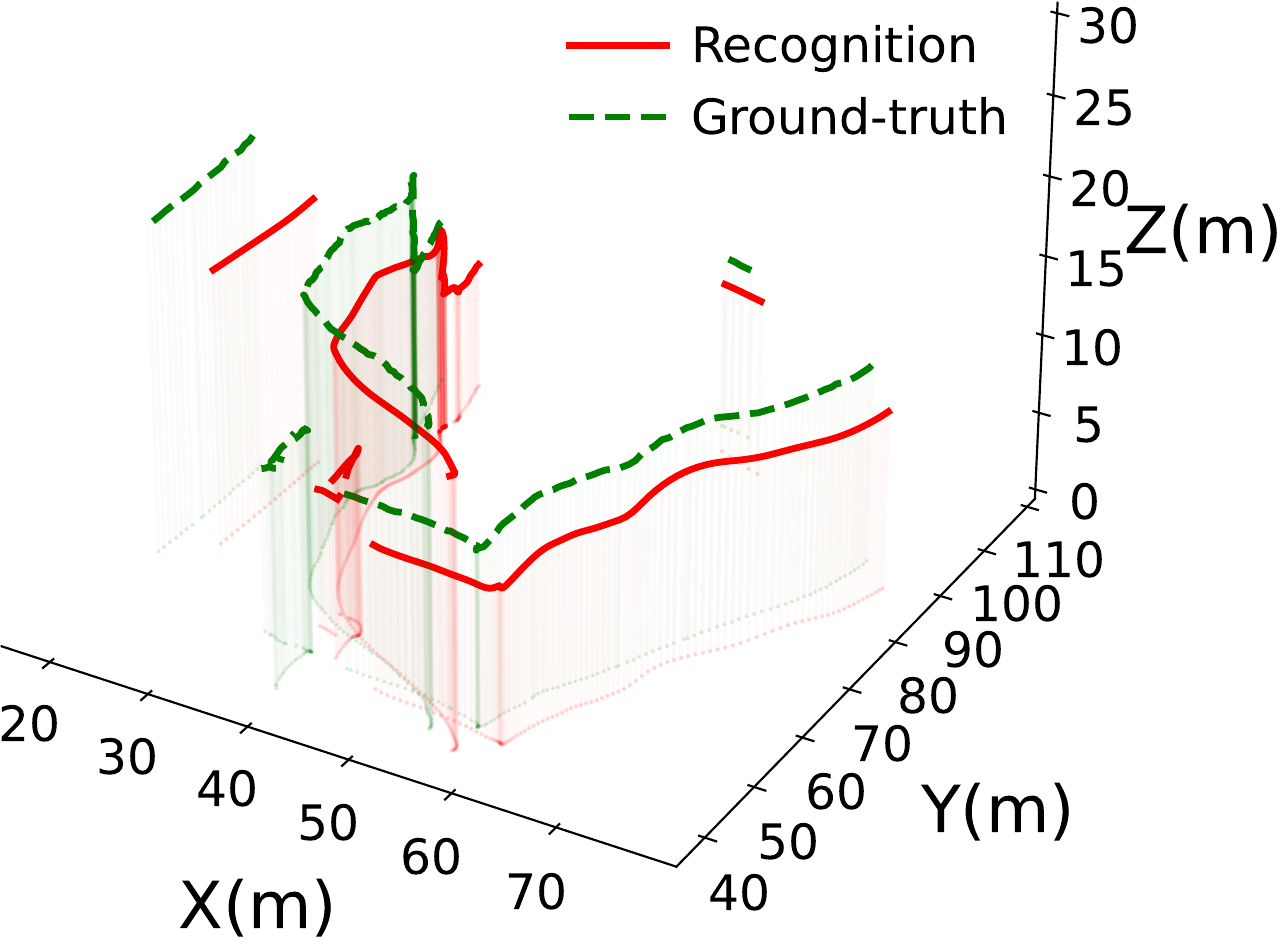}
\caption{
Visualization of the reconstructed UAV flight trajectory. The solid red line represents the UAV flight trajectory obtained from 3D coordinate recognition, connecting the discrete points. The dashed green line indicates the ground-truth UAV flight trajectory. The points on the ground are projections of these coordinates.
}
\label{fig:reconstructed-trajectory-result}
\end{figure*}

\begin{figure}[htbp]
\centering
\includegraphics[width=1\linewidth]{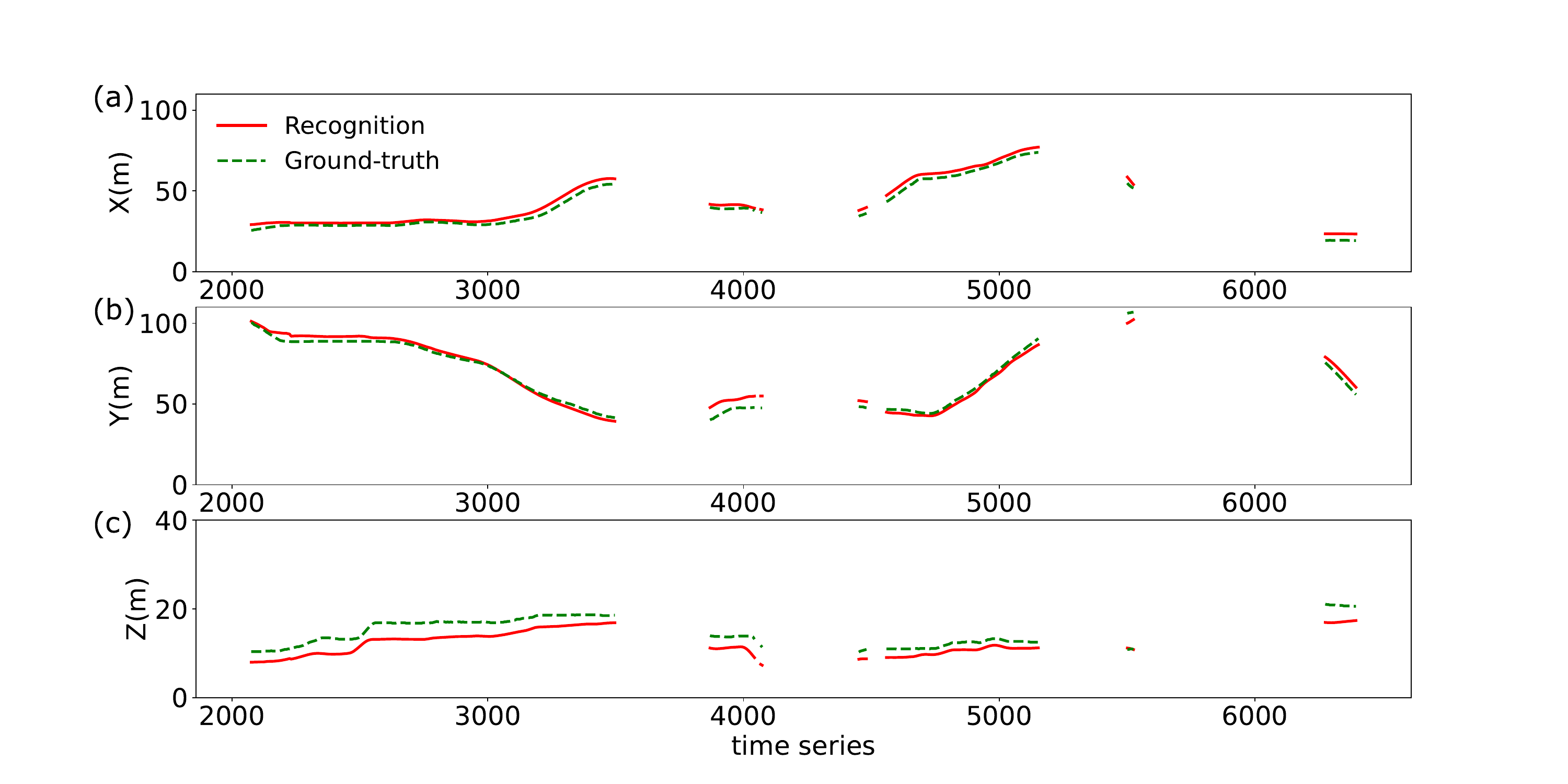}
\caption{
Evaluation of UAV 3D coordinate recognition results with ground-truth along the X, Y, and Z axes. The figure is divided into three sections: (a) shows the deviation of the UAV 3D coordinates along the X axis, (b) along the Y axis, and (c) along the Z axis. The solid red lines represent the UAV 3D coordinates obtained from the recognition process, while the dashed green lines indicate the corresponding ground-truth UAV 3D coordinates provided by the onboard positioning device.
}
\label{fig:reconstructed-trajectory-resultXYZ}
\end{figure}

The results of UAV coordinate recognition are depicted in \autoref{fig:reconstructed-trajectory-result}. \autoref{fig:reconstructed-trajectory-resultXYZ} presents the UAV coordinate recognition results in the world coordinate system, displayed across three graphs corresponding to the X, Y, and Z components of UAV 3D coordinates.

The metrics used in the UAV experiment include the overall and component-wise (X, Y, Z) RMSE, MAE, Maximum Error, and R-squared (Section~\ref{sec:metrics}). For the overall recognition, the RMSE is \(4.94357 \, \text{m}\), MAE is \(4.79475 \, \text{m}\), Maximum Error is \(8.84852 \, \text{m}\), and R-squared is \(0.95673\).

For the X-axis, the RMSE is \(2.77908 \, \text{m}\), MAE is \(2.59418 \, \text{m}\), Maximum Error is \(4.39648 \, \text{m}\), and R-squared is \(0.96459\), making it the most accurate among the three axes.
For the Y-axis, the RMSE is \(3.08915 \, \text{m}\), MAE is \(2.63404 \, \text{m}\), Maximum Error is \(8.13906 \, \text{m}\), and R-squared is \(0.97170\).
For the Z-axis, the RMSE is \(2.67820 \, \text{m}\), MAE is \(2.54572 \, \text{m}\), Maximum Error is \(4.26897 \, \text{m}\), and R-squared is \(0.24440\). 
The results of the UAV experiment indicate that our proposed time-series-based applied mathematical method for spatial point coordinate recognition is both robust and effective, achieving an RMSE of \(4.94357 \, \text{m}\) and an R-squared of \(0.95673\) in a \(100 \times 100 \times 30\, \text{m}\) scenario, without prior knowledge of camera attitudes.

% To quantitatively evaluate the performance of our method, we use the Root Mean Square Error (RMSE) between the computed spatial point coordinates and the ground truth spatial point coordinates as the evaluation metric (Section~\ref{sec:metrics}). 
% The average deviation of the UAV coordinates was 4.79 m, with a median error of 4.49 m. 
% rmse为，mae为，max偏移为

% The average error in the x-direction was 2.59 m, the average error in the y-direction was 0.82 m, and the average error in the z-direction was 2.54 m. The RMSE of the calculated coordinates compared to the ground truth coordinates was 4.94 m.

\section{Discussion}
Current methods for 3D positioning of non-cooperative targets in the world coordinate system typically depend on calibration objects or attitude measurement devices. 
We propose a framework for 3D coordinate recognition that removes the process of attitude measurement, 
incorporating AI-driven 2D detection technology and a mathematically-based 2D-to-3D transformation.
First, the core of the framework lies in applied mathematics. 
It uses time series and SVD to calculate the relative poses of cameras, offering an effective way to determine camera attitudes. 
Additionally, it employs an SVD-based method to calculate the similarity transformation matrix to derive the camera-to-world coordinate transformation.
Second, the framework optimizes AI by integrating the physical characteristics of moving objects into detection, significantly improving the accuracy of 2D coordinate time series acquisition.
Third, the framework serves a broader purpose in Earth system science, demonstrating versatility by integrating artificial intelligence and applied mathematics for geomatics theory.
Although the UAV is used as a case in our experiment, the method can be applied to other aerial objects, such as birds, and even to larger objects composed of distinct feature points.
Extending this method to multi-target scenarios, where 2D coordinate of multiple objects
are available, provides enhanced observations for calculating camera poses and potentially improves 3D coordinate recognition accuracy. 

Finally, our current method theoretically offers error-free results, however in engineering scenarios, the accuracy of 3D coordinate recognition is influenced by factors such as 2D coordinate deviations and camera positioning errors, particularly in larger scene sizes. 
These errors remain within acceptable limits for most applications, but further reductions must come from improved detection models and higher device precision. Notwithstanding the remaining limitations and future work, the proposed method contributes to geodetic theory by creating a more streamlined and efficient geodetic positioning approach, free from traditional attitude input parameters.

\section{Methods}
% \autoref{fig:UAV-experiment-schematic-diagram} is our method flow. In the coordinate recognition of 3D points, the accuracy of transforming coordinates between 2D camera pixel systems and 3D world systems is essential. Such transformations depend on a well-defined coordinate system model. Therefore, this section first introduces the coordinate systems involved, outlining their definitions and the transformations required.

\begin{figure}[htbp]
    \centering
    \includegraphics[width=\textwidth]{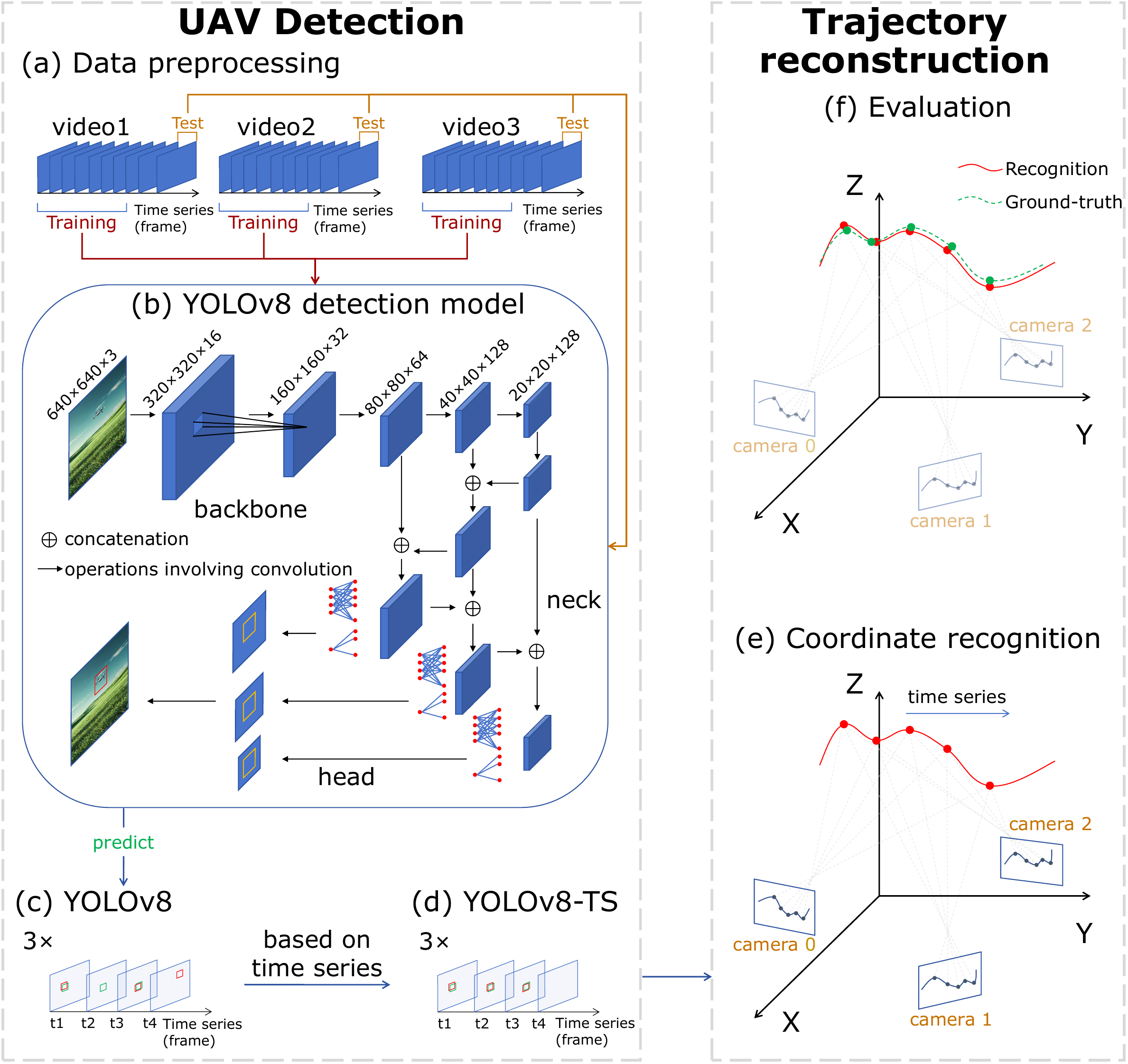}
    \caption{
    Schematic diagram of the UAV experiment for trajectory reconstruction.
    (a) Data Preprocessing: 
    Collecting the images from the UAV flight captured by three cameras and dividing them into training and test sets with a ratio of 8:2.
    (b) YOLOv8 Model Training: 
    Training the UAV detection model based on the YOLOv8 framework with a training set of UAV images captured in various scenes.
    (c) YOLOv8-Based UAV Prediction: 
    The trained model is used to predict the bounding boxes of the UAV in the videos captured by the three cameras.
    (d) YOLOv8-TS: 
    The predicted UAV trajectory in each video is further refined using our proposed YOLOv8-TS (see \autoref{sec:yolov8-TS} for details).
    (e) 3D Coordinate Recognition: 
    The UAV 2D coordinate time series are used for UAV 3D coordinate recognition using our proposed method (\autoref{sec:computation principle for world coordinate calculation}), and the flight trajectory is further reconstructed.
    (f) Evaluation: 
    The reconstructed trajectory is evaluated against ground-truth 3D coordinate data provided by the UAV onboard positioning device. The metrics used are RMSE, MAE, Maximum Error, and R-squared, as detailed in \ref{sec:metrics}.
    }
    \label{fig:UAV-experiment-schematic-diagram}
\end{figure}

\subsection{Formulation of the coordinate system model}
 Within a three-dimensional space, as shown in \autoref{fig:multiple-coordinate-system-transformation}, we consider a set of ground-based cameras, each uniquely represented by its optical center for camera \(i\), denoted as \( G^i \), where \( i = 0, 1, 2, \ldots, N-1 \), and a observed flying object denoted as \( O^j \) at each time step \( j \), where \( j = 0, 1, 2, \ldots, M-1 \). The field of view of these cameras covers the object \( O^j \) at all time steps.

 The coordinates of a spatial point \(P\) (\( G^i \) or \( O^j \)) differ depending on the chosen coordinate system. 
 In the trajectory reconstruction of flying object, the world coordinate system, camera coordinate system, and pixel coordinate system are typically employed. 
 Here, the trajectory of the flying object is first obtained in the pixel coordinate systems of each camera, which is then transformed into the camera coordinate systems. Next, the trajectory in the camera coordinate systems can be converted to the world coordinate system via the similarity transformation.
 The schematic of the our proposed methodology is shown in \autoref{fig:multiple-coordinate-system-transformation}.

\begin{enumerate}
  \item \textbf{World coordinate system}: 
  There are two types of world coordinate systems, the 3D spherical coordinate system and the 3D Cartesian coordinate system. 
  We adopt the latter to facilitate our subsequent mathematical modeling and linear algebra analysis.
  In this study, we use the World Geodetic System 1984 \cite{wgs84definition} reference frame. As shown in \autoref{fig:multiple-coordinate-system-transformation}, our 3D Cartesian coordinate system has its origin at the Earth center of mass. The Z-axis is aligned towards the Celestial Terrestrial Pole (CTP), as specified by the Bureau International de l'Heure (BIH 1984.0). The X-axis corresponds to the intersection of the BIH 1984.0 zero meridian and the equator of the CTP. The Y-axis, together with the Z-axis and X-axis, forms a right-handed coordinate system and can thus be directly determined. We denote the world coordinate system as \( w \). The coordinates of each point \( P \) in \( w \) are represented as \( P_{w} = (X_w, Y_w, Z_w) \).
  
  \item \textbf{Camera coordinate system}: 
 We define the camera coordinate system, which is also a 3D Cartesian coordinate system, with its origin located at the camera optical center in \autoref{fig:multiple-coordinate-system-transformation}(b). The Z-axis of the camera coordinate system is perpendicular to the camera imaging plane, while the X-axis and Y-axis are parallel to the horizontal and vertical directions of the imaging plane, respectively.
  We denote the camera coordinate system (\(c\)) for each camera \(i\) as ${c^i}$. The coordinates of each point \( P \) in \( c^i \) are represented as \( P_{c^i} = (X_{c^i}, Y_{c^i}, Z_{c^i}) \). 
  
  \item \textbf{Pixel coordinate system}:
  \autoref{fig:multiple-coordinate-system-transformation}(b) shows a point \( P \) in a 3D space appears as a pixel after being imaged by a camera. The captured image can be represented by a matrix, where the row and column indices of a pixel in this matrix correspond to its coordinates in the 2D pixel coordinate system.
  The origin of this coordinate system is located at the top-left corner of the image, with the \( u \)-axis pointing to the right and the \( v \)-axis pointing downwards.
  We denote the pixel coordinate system (\(c'\)) for each camera \(i\) as \({c'}^i\). 
  The coordinates of each point \(P\) in \({c'}^i\) are represented as \(P_{{c'}^i} = (u_{{c'}^i}, v_{{c'}^i})\). For computational convenience, this 2D coordinate is extended to a 3D coordinate \( (u_{{c'}^i}, v_{{c'}^i}, 1) \), which is referred to as the homogeneous pixel coordinate. In this paper, we use \( (u_{{c'}^i}, v_{{c'}^i}, 1) \) as our pixel coordinates.

\end{enumerate}

\subsection{Coordinate transformations}
\begin{enumerate}
  \item \textbf{3D to 3D}: 
  Based on a similarity transformation that includes a rotation matrix, a translation vector, and a scale factor, the coordinates of a point \( P \) can be transformed from one 3D Cartesian coordinate systems \( m \) to another \( n \):
  \begin{equation}
  P_{n} = s_{nm} R_{nm} P_{m} + t_{nm} \label{eq:transf3d},
  \end{equation}
  where \( P_{n} \) denotes the coordinates of \( P \) in the coordinate system \( n \), \( P_{m} \) denotes the coordinates of \( P \) in the coordinate system \( m \), \( s_{nm} \) is a constant scale factor that represents the scaling from \( m \) to \( n \), \( R_{nm} \) is a \( 3 \times 3 \) rotation matrix from \( m \) to \( n \), and \( t_{nm} \) is a \( 3 \times 1 \) translation vector in \( n \) pointing from the origin of \( n \) to the origin of \( m \).
  In particular, if the coordinate systems \( n \) and \( m \) are at the same scale, then \( s_{nm} = 1 \). For the world coordinate systems \( w \) and the camera coordinate system \( c \), we have:
  \begin{equation}
  P_{c} = {R_{cw}} P_{w} +{t_{cw}} \label{eq:3d23d}.
  \end{equation}

  \item \textbf{3D to 2D}:
  The point \( P \) in the 3D camera coordinate system is transformed into the 2D pixel coordinate system based on the pinhole model (\autoref{fig:multiple-coordinate-system-transformation}(b)), with this transformation process precisely described by the camera intrinsic parameter matrix \( K \):
  \[
  K = \begin{bmatrix}
  f_x & 0 & d_x \\
  0 & f_y & d_y \\
  0 & 0 & 1
  \end{bmatrix},
  \]
  where \( f_x \) and \( f_y \) are the focal lengths in the x and y directions, respectively, and \( (d_x, d_y) \) is the principal point, which is the projection of the camera optical center onto the image plane.

  Given a point \( P = (X, Y, Z) \) in the camera coordinate system, the projected pixel coordinates \( (u, v, 1) \) can be obtained by:
  \begin{equation}
Z\begin{bmatrix}
  u \\
  v \\
  1
  \end{bmatrix} = K \begin{bmatrix}
  X \\
  Y \\
  Z
  \end{bmatrix} = \begin{bmatrix}
  f_x & 0 & d_x \\
  0 & f_y & d_y \\
  0 & 0 & 1
  \end{bmatrix} \begin{bmatrix}
  X \\
  Y \\
  Z
  \end{bmatrix}
  \label{eq:3d22d}.
  \end{equation}
  
  \item \textbf{2D to 3D}:
  \autoref{eq:3d22d} maps a point \( P \) from a 3D coordinate system to a 2D coordinate system, however this mapping is non-invertible due to the ambiguity in the depth dimension \( Z \). Resolving depth \( Z \) relies on identifying corresponding points across images from multiple cameras to establish spatial relationships.
  
  For two camera coordinate systems \( {c^1} \) and \( {c^2} \), given the rotation matrix \( R_{c^2c^1} \) and the translation vector \( t_{c^2c^1} \), the coordinates of point \( P \) can be transformed between \( {c^1} \) and \( {c^2} \) according to \autoref{eq:transf3d}:
  \begin{equation}
  P_{c^2} = R_{c^2c^1} P_{c^1} + t_{c^2c^1}
  \label{eq:c12c2}.
  \end{equation}

  Given the pixel coordinates \( P_{{c'}^1} \) and \( P_{{c'}^2} \) of point \( P \) in the pixel coordinate systems \({c'}^1\) and \({c'}^2\), we can obtain the coordinate transformations for point \( P \) from \(c^1\) to \({c'}^1\) and from \(c^2\) to \({c'}^2\) based on \autoref{eq:3d22d}:
  \begin{equation}
  Z_{c^1} P_{{{c'}^1}} = K^{1} P_{c^1}
  \label{eq:pinholep1},
  \end{equation}
  \begin{equation}
  Z_{c^2} P_{{{c'}^2}} = K^{2} P_{c^2}
  \label{eq:pinholep2},
  \end{equation}
  where \(Z_{c^1}\) and \(Z_{c^2}\) represent the depth Z of point \(P\) in \({c^1}\) and \({c^2}\), respectively.
  Substituting \autoref{eq:pinholep1} and \autoref{eq:pinholep2} into \autoref{eq:c12c2}, we obtain
  \begin{equation}
    Z_{c^2} P_{{{c'}^2}} = Z_{c^1} {K^{2}} R_{c^2c^1} {K^{1}}^{\text{-}1} P_{{{c'}^1}} + {K^{2}} t_{c^2c^1}.
    \label{eq:beforeskew}
  \end{equation}
  Multiplying both sides of \autoref{eq:beforeskew} on the left by the matrix \([ P_{{{c'}^2}} ]_\times\), we obtain:
  \begin{equation}
  Z_{c^2} [P_{{{c'}^2}}]_\times P_{{{c'}^2}} = Z_{c^1} [P_{{{c'}^2}}]_\times {K^{2}} R_{c^2c^1} {K^{1}}^{\text{-}1} P_{{{c'}^1}} + [P_{{{c'}^2}}]_\times {K^{2} t_{c^2c^1}} 
  \label{eq:multiplyingskew-symmetricmatrix},
  \end{equation}

  where \([ P_{{{c'}^2}} ]_\times\) is the skew-symmetric matrix of \( P_{{{c'}^2}} \). For vector \( P_{{{c'}^2}} = [x_{{{c'}^2}}, y_{{{c'}^2}}, z_{{{c'}^2}} ]^T \), its corresponding skew-symmetric matrix \( [P_{{{c'}^2}}]_\times \) is defined as follows:
  \begin{equation}
  [P_{{{c'}^2}}]_\times = \begin{bmatrix}
  0 & \text{-}z_{{{c'}^2}} & y_{{{c'}^2}} \\
  z_{{{c'}^2}} & 0 & \text{-}x_{{{c'}^2}} \\
  \text{-}y_{{{c'}^2}} & x_{{{c'}^2}} & 0 
  \end{bmatrix}
  \label{eq:skew-symmetricmatrix}.
  \end{equation}
  From \autoref{eq:skew-symmetricmatrix}, we can clearly conclude that the product \([P_{{{c'}^2}}]_\times P_{{{c'}^2}}\) of a skew-symmetric matrix \([P_{{{c'}^2}}]_\times\) and its corresponding vector \(P_{{{c'}^2}}\) results in a zero vector. Therefore, the left-hand side of \autoref{eq:multiplyingskew-symmetricmatrix} equals the zero vector and we get

  \begin{equation}
  Z_{c^1} = \frac{\text{-} [P_{{{c'}^2}}]_\times {K^{2}} R_{c^2c^1} {K^{1}}^{\text{-}1} P_{{{c'}^1}} \cdot [P_{{{c'}^2}}]_\times {K^{2} t_{c^2c^1}}}{[P_{{{c'}^2}}]_\times {K^{2}} R_{c^2c^1} {K^{1}}^{\text{-}1} P_{{{c'}^1}} \cdot [P_{{{c'}^2}}]_\times {K^{2}} R_{c^2c^1} {K^{1}}^{\text{-}1} P_{{{c'}^1}}}.
  \end{equation}

  Substituting the above solution \(Z_{c^1}\) into \autoref{eq:pinholep1}, we obtain the coordinates of \(P\) in \(c^1\):
  \begin{equation}
    P_{c^1} = \frac{\text{-}[P_{{{c'}^2}}]_\times {K^{2}} R_{c^2c^1} {K^{1}}^{\text{-}1} P_{{{c'}^1}} \cdot [P_{{{c'}^2}}]_\times {K^{2} t_{c^2c^1}}}{[P_{{{c'}^2}}]_\times {K^{2}} R_{c^2c^1} {K^{1}}^{\text{-}1} P_{{{c'}^1}} \cdot [P_{{{c'}^2}}]_\times {K^{2}} R_{c^2c^1} {K^{1}}^{\text{-}1} P_{{{c'}^1}}} {K^{1}}^{\text{-}1} P_{{c'}^1}
    \label{eq:finalstep2d23d}.
  \end{equation}
\end{enumerate}

\subsection{
Method of coordinate recognition
\label{sec:computation principle for world coordinate calculation}
}

Under the condition where only the world coordinates of multiple ground-based observation cameras \( G_{w}^i \) (where \(i \geq 3, \, i \in \mathbb{N} \)) are available, our proposed method does not require knowledge of the camera attitude \(R_{c^iw}\), which is necessary for  conventional methods \cite{sie2021field} to calculate the world coordinates of \( O^j \).

Given the coordinates (\( O_{{c'}^0}^j \) and \( O_{{c'}^1}^j \)) of spatial point \( O^j \) in the pixel coordinate systems \( {c'}^0 \) and \( {c'}^1 \), respectively, the relationship between \( O_{{c'}^0}^j \) and \( O_{{c'}^1}^j \) can be derived based on the principles of epipolar geometry \cite{longuet1981computer}, as follows:

\begin{equation}
    {O_{{c'}^1}^j}^T F_{c^1c^0} O_{{c'}^0}^j = 0.
    \label{eq:projective-geometry}
\end{equation}

To solve for the fundamental matrix \( F_{c^1c^0} \), which describes the epipolar geometry between two views, a minimum of 7 pairs of corresponding points (\( O_{{c'}^0}^j \) and \( O_{{c'}^1}^j \)) is required. However, 8 pairs are typically used to ensure a more stable and unique solution, especially when using the RANSAC algorithm to robustly estimate \( F_{c^1c^0} \) in the presence of outliers. The fundamental matrix \( F_{c^1c^0} \) is defined as:

\begin{equation}
F_{c^1c^0} ={K^1}^{\text{-}T} E_{c^1c^0} {K^0}^{\text{-}1},
\end{equation}
where \( K^0 \) and \( K^1 \) are the intrinsic parameter matrices of camera \( 0 \) and \( 1 \), respectively. The essential matrix \( E_{c^1c^0} \), which relates corresponding points between two calibrated views and encodes the relative rotation and translation from \(c^1\) to \(c^0\), is given by:

\begin{equation}
\label{eq:essential-matrix}
E_{c^1c^0} = [t_{c^1c^0}]_\times R_{c^1c^0}.
\end{equation}

In this equation, \( [t_{c^1c^0}]_\times \) is the skew-symmetric matrix of the translation vector \( t_{c^1c^0} \). The essential matrix \( E_{c^1c^0} \) can be decomposed using singular value decomposition (SVD) and  we define the extracted \( t_{c^1c^0} \) as \( \tilde{t}_{c^1c^0} \) and the extracted \( R_{c^1c^0} \) as \( \tilde{R}_{c^1c^0} \).

Given that multiplying \( F_{c^1c^0} \) by any constant still satisfies \autoref{eq:projective-geometry}, there is a scale ambiguity in both \( F_{c^1c^0} \) and \( E_{c^1c^0} \).  For a rotation matrix \( R \), given its inherent property \( \det(R) = 1 \), we have \( \tilde{R}_{c^1c^0} = R_{c^1c^0} \). For \( \tilde{t}_{c^1c^0} \), there exists a scale factor \( s \) such that 

\begin{equation}
    \tilde{t}_{c^1c^0} = s t_{c^1c^0}.
    \label{eq:extractedt}
\end{equation}

From \autoref{eq:transf3d}, the coordinates of a point \( P \) from \( c^0 \) to \( c^1 \) can be transformed by

\begin{equation}
P_{c^1} = R_{c^1c^0} P_{c^0} + t_{c^1c^0}.
\label{eq:c1c0}
\end{equation}

Substituting \(R_{c^1c^0}\) and \(t_{c^1c^0}\) into \autoref{eq:c1c0}, we have:
\begin{equation}
    s P_{c^1} = s \tilde{R}_{c^1c^0} P_{c^0} + \tilde{t}_{c^1c^0}.
    \label{eq:sPc1}
\end{equation}

Now, we define a scaled coordinate system \( \hat{c}^i \) for each camera coordinate system \( c^i \) with a scale factor \(s\). The coordinates of a point \( P \) from \( c^i \) to \( \hat{c}^i \) are transformed by:
\begin{equation}
P_{\hat{c}^i} = s P_{c^i}.
\label{eq:chat}
\end{equation}

Thus, Eq~\ref{eq:sPc1} can be rewritten as:
\begin{equation}
P_{\hat{c}^1} = \tilde{R}_{c^1c^0} P_{\hat{c}^0} + \tilde{t}_{c^1c^0}.
\end{equation}

Therefore, based on the definition of similarity transformation between coordinates in different coordinate systems,
\begin{equation}
R_{\hat{c}^1\hat{c}^0} = \tilde{R}_{c^1c^0} = R_{c^1c^0},
\label{eq:Rhatc1hatc0}
\end{equation}

\begin{equation}
t_{\hat{c}^1\hat{c}^0} = \tilde{t}_{c^1c^0} = s t_{c^1c^0}.
\label{eq:thatc1hatc0}
\end{equation}

From \autoref{eq:finalstep2d23d}, we can derive the expression for \(O^j_{c^0}\):
\begin{equation}
    O^j_{c^0} = - \frac{[O_{{c'}^1}]_\times {K^{1}} R_{c^1c^0} {K^{0}}^{\text{-}1} O_{{c'}^0} \cdot [O_{{c'}^1}]_\times {K^{1}} t_{c^1c^0}} {[O_{{c'}^1}]_\times {K^{1}} R_{c^1c^0} {K^{0}}^{\text{-}1} O_{{c'}^0} \cdot [O_{{c'}^1}]_\times {K^{1}} R_{c^1c^0} {K^{0}}^{\text{-}1} O_{{c'}^0}} {K^{0}}^{\text{-}1} O^j_{{c'}^0}.
    \label{eq:Oc0j}
\end{equation}

Substituting \autoref{eq:chat}, \autoref{eq:Rhatc1hatc0} and \autoref{eq:thatc1hatc0} into \autoref{eq:Oc0j}, we obtain
\begin{equation}
\label{eq:triangulation-O}
    O^j_{\hat{c}^0} = \text{-} \frac{{O_{{c'}^0}^T [O_{{c'}^1}]_\times {K^{1}} \tilde{R}_{c^1c^0} {K^{0}}^{\text{-}1} O_{{c'}^0}} \cdot {O_{{c'}^0}^T[O_{{c'}^1}]_\times {K^{1}} \tilde{t}_{c^1c^0}}} {{O_{{c'}^0}^T [O_{{c'}^1}]_\times {K^{1}} \tilde{R}_{c^1c^0} {K^{0}}^{\text{-}1} O_{{c'}^0}} \cdot {O_{{c'}^0}^T [O_{{c'}^1}]_\times {K^{1}} \tilde{R}_{c^1c^0} {K^{0}}^{\text{-}1} O_{{c'}^0}}} {K^{0}}^{\text{-}1} O^j_{{c'}^0}.
\end{equation}
Given the coordinates \(O_{\hat{c}^0}^j\) of the spatial observed points \(O^j\) in \(\hat{c}^0\) and coordinates \({O}_{{c'}^i}^j\) in the pixel coordinate systems \( {c'}^i \) of cameras \(i = 2, 3, ..., N\), the rotation matrix \( R_{\hat{c}^i\hat{c}^0} \) and translation vector \( t_{\hat{c}^i\hat{c}^0} \) can be determined based on the Efficient Perspective-n-Point (EPnP) algorithm \cite{lepetit2009ep}. The EPnP method represents these points as a linear combination of four virtual control points and uses SVD to solve these linear equations.

Next, we calculate ${ R_{w\hat{c}^0} }$ and ${t_{w\hat{c}^0}}$. From \autoref{eq:3d23d} and \autoref{eq:chat}, we can derive the coordinate transformation of a spatial point \(P\) from \(w\) to \(\hat{c}^0\):

\begin{equation}
P_{\hat{c}^0} = s R_{c^0w}P_w+ t_{\hat{c}^0w}.
\label{eq:Pchat0w}
\end{equation}

For the cameras, \autoref{eq:Pchat0w} can be rewritten:
\begin{equation}
G_{\hat{c}^0}^i = s R_{c^0w}G_w^i+ t_{\hat{c}^0w}.
\label{eq:Gchat0w}
\end{equation}

Here, \(G_w^i\) is known. \(G_{\hat{c}^0}^i\) represents the coordinates of camera \(i\) in the coordinate system \( \hat{c}^0 \), and \(t_{\hat{c}^0\hat{c}^i}\) as well. Based on the inverse operation of \autoref{eq:transf3d}, \(t_{\hat{c}^0\hat{c}^i}\) can be calculated by:
\begin{equation}
\label{eq:camera-coordinate-in-camera0}
G_{\hat{c}^0}^i = t_{\hat{c}^0\hat{c}^i} =  - {R_{\hat{c}^i\hat{c}^0}}^{\text{-}1} t_{\hat{c}^i\hat{c}^0}.
\end{equation}

We use the Kabsch algorithm \cite{kabsch1978discussion} for solving the similarity transformation to determine \(s\), \(R_{\hat{c}^0w}\), and \(t_{\hat{c}^0w}\) in \autoref{eq:Gchat0w}. First we calculate the centroid \(\overset{\circ}{G_{w}}\) and \( \overset{\circ}{G_{\hat{c}_0}} \) for \( G_{w}^{i} \) and \( G_{\hat{c}_0}^{i}\) :

\[ 
\overset{\circ}{G_{w}} = \frac{1}{N} \sum_{i=0}^{N-1} G_{w}^{i},
\]

\[
\overset{\circ} {G_{\hat{c}_0}} = \frac{1}{N} \sum_{i=0}^{N-1} G_{\hat{c}_0}^{i}.
\]

Let
\[\overset{*}{{G}_{w}^{i}} = {G}_{w}^{i} - \overset{\circ}{G_{w}},\]
\[\overset{*}{{G}_{\hat{c}_0}^{i}} = {G}_{\hat{c}_0}^{i} - \overset{\circ}{G_{\hat{c}_0}},\]

\[
H = \sum_{i=0}^{N-1} \overset{*}{{G}_{\hat{c}_0}^{i}} ( \overset{*}{{G}_{w}^{i}} )^{T}.
\]
Find the SVD of $H$

\[H = U \Lambda V^{T},\]

Then, \(R_{\hat{c}^0w}\), \(s\) and \(t_{\hat{c}^0w}\) can be computed as follows:
\[ R_{\hat{c}^0w} = VU^{T}, \]

\[
s = \sqrt{\frac{\sum_{i=0}^{N} \|\overset{*}{{G}_{\hat{c}_0}^{i}}\|^2}{\sum_{i=0}^{N} \|\overset{*}{{G}_{w}^{i}}\|^2}},
\]

\[
t_{\hat{c}^0w} = G_{\hat{c}^0}^i - s R_{c^0w}G_w^i.
\]

Now we recall \autoref{eq:Pchat0w}, whose inverse process transforms the coordinates of point \( P \) from \(\hat{c}^0\) to \(w\):
\begin{equation}
P_w^j=\frac{{R_{\hat{c}^0w}}^{\text{-}1} ( P_{\hat{c}^0}^j - t_{\hat{c}^0w} )}{s}.
\end{equation}

Given \(s\), \(R_{\hat{c}^0w}\), and \(t_{\hat{c}^0w}\), the coordinates \(O_w^j\) of the observed spatial object \(O^j\) in the world coordinate system (\(w\)) at the \(j\)-th moment in the time series from 0 to \(M-1\) are obtained by:

\begin{equation}
\label{eq:object-coordinate-world}
O_w^j=\frac{{R_{\hat{c}^0w}}^{\text{-}1} ( O_{\hat{c}^0}^j - t_{\hat{c}^0w} )}{s}.
\end{equation}

\subsection{YOLOv8-Time series 
\label{sec:yolov8-TS}
}
In YOLOv8-based UAV detection, we face two main challenges. First, other moving objects, such as birds, can cause interference. 
Second, missed detections can occur due to low confidence levels or complex backgrounds involving buildings. However, the unique time series of UAV flight and its speed-based physical characteristics can be extracted to distinguish it from other small moving objects and mitigate missed detections. Based on this, we propose YOLOv8-Time Series (YOLOv8-TS).
After using the trained YOLOv8 model for UAV prediction (\autoref{fig:UAV-experiment-schematic-diagram}(c)), we employ the following methods to further confirm whether the detected object is a UAV or something else, and to supplement missed detections.
Each detection result haves one of two statuses: confirmed as a UAV, or pending confirmation. 
We apply the following rules to identify all confirmed UAV detections, where Rule 1 is mainly used for UAV confirmation, and Rule 2 is primarily used for supplementing missed UAV detections.

\textbf{Rule 1.}  
For the \(i\)-th frame, if both the \(i\)-th and \(j\)-th frames have detection results and the distance \(d(i, j)\) between them is less than \(10 \times (j-i)\) pixels, we consider the detection results of both frames as confirmed UAVs.

\textbf{Rule 2.}  
Based on the confirmed UAV segments obtained from Rule 1, we calculate the UAV speed between adjacent frames as the distance between the UAV positions. 
From these calculated speed, we determine the maximum non-outlier speed. 
For the \(i\)-th frame and the \(i+j\)-th frame \((1 < j \leq 180, j \in \mathbb{Z})\), if both frames have confirmed UAV detections, the center coordinates \(x\) and \(y\) of the detections are within 1\% to 99\% of the image width and height, and the distance between the \(i\)-th and \(i+j\)-th frames is less than \(j \times\) the maximum non-outlier speed, or if the interval between the frames is less than 60 frames, we determine that there is an undetected UAV in the frames between the \(i\)-th and \(i+j\)-th frames. For each \(i+m\)-th frame \((1 < m < j, m \in \mathbb{Z})\), the coordinates and dimensions are interpolated between the \(i\)-th and \(i+j\)-th frames, and the status of the \(i+m\)-th frame is set to confirmed UAV.

\begin{algorithm}
\caption{YOLOv8-Time series}
\begin{algorithmic}[1]
\Require \textbf{frames}: time series of frames with detection results

\Statex
\For{each frame $i$ in $frames$}
    \For{$j = i+1$ to $i+3$}
        \If{$\text{distance}(i, j) < 10 \times (j - i)$}
            \State Mark $detection\_result[i]$ and $detection\_result[j]$ as confirmed UAVs
        \EndIf
    \EndFor
\EndFor

\Statex

\State $UAV\_speed \gets$ Distances between consecutive confirmed UAV frames
\State $max\_non\_outlier\_speed \gets 2.5 \times \text{calculate\_max\_non\_outlier\_speed}(comfirmed\_UAV\_speed)$

\Statex

\For{each frame $i$ in $frames$ where $frames[i]$ is confirmed}
    \State $j \gets i + 1$
    \While{$j < \text{length of } frames$ and $frames[j]$ is not confirmed UAV}
        \State $j \gets j + 1$
    \EndWhile
    
    \State $time\_length \gets j - i$
    \State $coordinates\_in\_range \gets \text{check if } i \text{ and } j \text{ are within boundaries}$
    \If{($coordinates\_in\_range$ and $time\_length \leq 180$) or $time\_length \leq 60$}
        \If{$\text{distance}(i, j) < time\_length \times max\_non\_outlier\_speed$}
            \For{$m \gets 1$ to $time\_length - 1$}
                \State Interpolate coordinates and dimensions between frames $i$ and $j$
                \State Mark frame $i+m$ as confirmed UAV
            \EndFor
        \EndIf
    \EndIf
\EndFor
\Statex

\State \Return $frames$ with confirmed UAV markings

\end{algorithmic}
\end{algorithm}

\subsection{Metrics
\label{sec:metrics}
}

\textbf{Metrcis for object detection model.}
To comprehensively evaluate the performance of the object detection model, we introduce a set of metrics that consider different IoU thresholds for Precision (P), Recall (R), F1-score (F1), and Area Under the Curve (AUC). Before defining these metrics, we first clarify the basic concepts of True Positives (TP), False Positives (FP), True Negatives (TN), and False Negatives (FN):

True Positives (TP\(_{\text{IoU}}\)): The number of instances where the model correctly predicts an object and the predicted bounding box has an IoU with the ground truth bounding box that exceeds a specified IoU threshold. 

False Positives (FP\(_{\text{IoU}}\)): The number of instances where the model incorrectly predicts an object where there is none, or the predicted bounding box does not meet the IoU threshold with any ground truth bounding box.

True Negatives (TN\(_{\text{IoU}}\)): The number of frames where the model correctly identifies the absence of objects.

False Negatives (FN\(_{\text{IoU}}\)): The number of instances where the model fails to detect objects that are actually present.

It should be noted that in cases where there is a UAV in the frame, but the detected bounding box has an IoU with the ground truth bounding box that is less than the specified IoU threshold, this instance is considered both a False Positive (FP) and a False Negative (FN).

With these definitions in place, we can now define the evaluation metrics:

IoU-Weighted Precision (IoU-P): Average Precision across IoU thresholds at every 5\% increment from 50\% to 95\%:

\begin{equation}
\text{IoU-P} = \frac{1}{10} \sum_{k=0}^{9} P_{50 + 5k},
\end{equation}

where \textbf{\(Precision_{IoU} (P_{\text{IoU}})\)} is the proportion of correctly identified UAVs among all detections:

\begin{equation}
P_{\text{IoU}} = \frac{\text{TP}_{\text{IoU}}}{\text{TP}_{\text{IoU}} + \text{FP}_{\text{IoU}}}.
\end{equation}

IoU-Weighted Recall (IoU-R): Average Recall across IoU thresholds at every 5\% increment from 50\% to 95\%:

\begin{equation}
\text{IoU-R} = \frac{1}{10} \sum_{k=0}^{9} R_{50 + 5k},
\end{equation}

where \textbf{\(Recall_{IoU} (R_{\text{IoU}})\)} is the proportion of correctly identified UAVs among all actual UAV samples:

\begin{equation}
R_{\text{IoU}} = \frac{\text{TP}_{\text{IoU}}}{\text{TP}_{\text{IoU}} + \text{FN}_{\text{IoU}}}.
\end{equation}

IoU-Weighted F1-Score (IoU-F1): Average F1-score across IoU thresholds at every 5\% increment from 50\% to 95\%:

\begin{equation}
\text{IoU-F1} = \frac{1}{10} \sum_{k=0}^{9} F1_{50 + 5k},
\end{equation}

where \textbf{\(F1-Score_{IoU} (F1_{\text{IoU}})\)} is the harmonic mean of Precision and Recall:

\begin{equation}
F1_{\text{IoU}} = 2 \cdot \frac{P_{\text{IoU}} \cdot R_{\text{IoU}}}{P_{\text{IoU}} + R_{\text{IoU}}}.
\end{equation}

Area under the curve (AUC): The AUC is calculated by plotting the True Positive Rate (TPR) against the False Positive Rate (FPR) across different IoU thresholds, and then computing the area under this curve:

\begin{equation}
\text{AUC} = \int \text{TPR}(\text{FPR}) \, d(\text{FPR}),
\end{equation}

where TPR and FPR are calculated as follows:

\begin{equation}
\text{TPR}_{\text{IoU}} = \frac{\text{TP}_{\text{IoU}}}{\text{TP}_{\text{IoU}} + \text{FN}_{\text{IoU}}},
\end{equation}

\begin{equation}
\text{FPR}_{\text{IoU}} = \frac{\text{FP}_{\text{IoU}}}{\text{FP}_{\text{IoU}} + \text{TN}_{\text{IoU}}}.
\end{equation}

\textbf{Metrcis for 3D coordinate recognition.}
To evaluate the accuracy of coordinate recognition for the flying object, we employ severalmetrics:

Root Mean Square Error (RMSE):
\begin{equation}
    \text{RMSE} = \sqrt{\frac{\sum_{j=0}^{M-1} \left\| \tilde{O}_w^j - O_w^j \right\|_2^2}{M}}.
\end{equation}
    
Mean Absolute Error (MAE):
\begin{equation}
    \text{MAE} = \frac{1}{M} \sum_{j=0}^{M-1} \left\| \tilde{O}_w^j - O_w^j \right\|_1.
\end{equation}
    
Maximum Error:
\begin{equation}
    \text{Max Error} = \max_{j=0}^{M-1} \left\| \tilde{O}_w^j - O_w^j \right\|_2.
\end{equation}
    
R-squared (Coefficient of Determination):
\begin{equation}
    R^2 = 1 - \frac{\sum_{j=0}^{M-1} \left\| \tilde{O}_w^j - O_w^j \right\|_2^2}{\sum_{j=0}^{M-1} \left\| O_w^j - \bar{O}_w \right\|_2^2}.
\end{equation}

Here, \(\tilde{O}_w^j\) and \(O_w^j\) represent the computed and ground truth spatial point coordinates in the world coordinate system at time step \(j\), respectively, and \(M\) denotes the total number of time steps.

\section{Supplementary information}

\subsection{Parameter selection of numerical simulations
\label{sec:Parameter-selection-of-simulation-experiment}
}

In the numerical simulations, the camera positioning error levels are set to \(0.1, 0.3, 0.5, 0.7, 0.9, 1.1\,\text{m}\). In engineering scenarios, cameras use high-precision Real-Time Kinematic (RTK) GNSS devices for positioning, which involve two main sources of error: a \(10\,\text{cm}\) installation error between the GNSS receiver and the camera, and the GNSS receiver positioning error, which is less than \(10\,\text{cm}\) \cite{alkan2020comparative}.

The pixel deviation levels in the numerical simulations are set to \(1, 2, 4, 6, 8, 10\) pixels. 
Consider a standard camera with a resolution of \(1920 \times 1080\) pixels, no geometric distortion, and an optical axis centered in the image, with focal lengths of \(1000\) in both the \(x\) and \(y\) directions. For a sphere with a diameter of \(0.5\,\text{m}\) located along the camera principal axis, the sphere occupies \(20\) pixels when it is \(25\,\text{m}\) away, \(10\) pixels at \(50\,\text{m}\), and \(2\) pixels at \(250\,\text{m}\) in both the \(x\) and \(y\) directions. The 2D coordinates of this target in the image are defined by the center of the occupied pixels. The deviation between any pixel on the occupied area and the center pixel position is half the length of the occupied pixels and is varied between \(1\) and \(10\) pixels.

The scale factors for the scene in the numerical simulations are set to \(0.25, 0.5, 1, 1.5,\) and \(2\), scaling the original scene size of \(200 \times 200 \times 100\,\text{m}\) to a range from \(50 \times 50 \times 25\,\text{m}\) to \(400 \times 400 \times 200\,\text{m}\).
Here, we set a fixed camera positioning error of \(0.2\,\text{m}\) and a pixel deviation of \(3\) pixels. 
For static camera positioning, we consider the camera positioning error is \(0.2\,\text{m}\), which includes a \(10\,\text{cm}\) positioning error from the high-precision RTK GNSS receiver and an additional \(10\,\text{cm}\) installation error.
For pixel deviations, we consider a \(0.5\,\text{m}\) spherical object. When it is positioned along the camera’s principal axis at a distance of \(83.3\,\text{m}\) from the optical center, it occupies \(6\) pixels in both the \(x\) and \(y\) directions, with a maximum pixel deviation of 3 pixels. While the number of pixels occupied by the object would vary at different distances, we simplify the experimental parameters by setting the pixel deviations to a fixed value of 3 pixels.

\subsection{
YOLOv8 Model
\label{sec:YOLOv8-structure}
}

\begin{figure}[htbp]
    \centering
    \includegraphics[width=\linewidth]{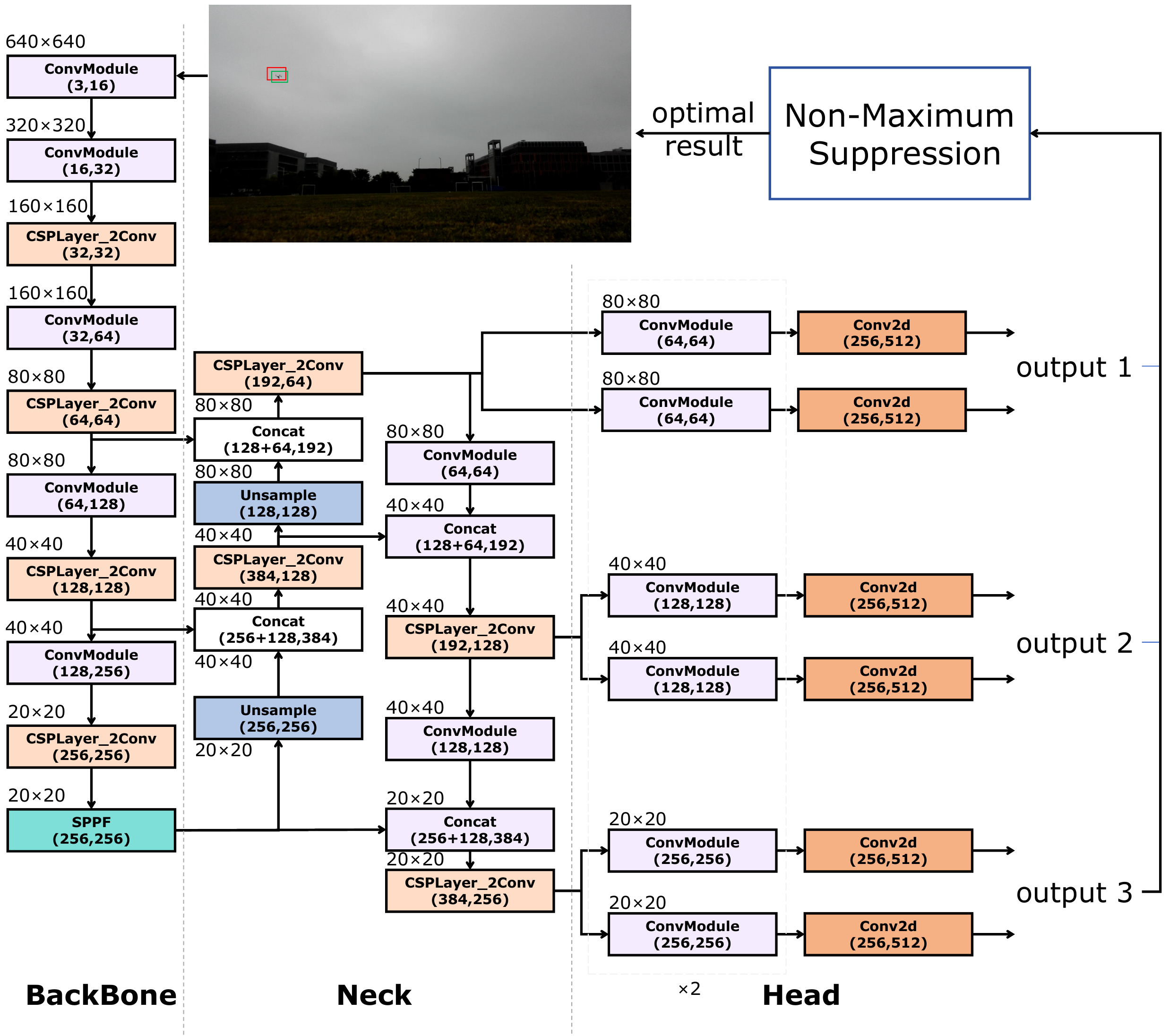}
    \caption{
    Overview of the network architecture of Yolov8.
    The parameters described here correspond to the YOLOv8n version used in this experiment.
    (a) Backbone: 
    The backbone is the core component responsible for extracting features from the input image. YOLOv8 utilizes a modified CSPDarknet53, a variant of the Darknet architecture. CSPDarknet53 integrates Cross Stage Partial Networks (CSPNet) to enhance gradient flow and reduce computational cost. CSPNet divides the input feature map into two parts: one part is processed through dense connections, while the other part is later concatenated back to improve gradient flow and learning capability. The backbone consists of several convolutional layers and residual blocks, which capture different levels of feature representations.
    (b) Neck: 
    The neck of YOLOv8 employs the Path Aggregation Feature Pyramid Network (PAFPN), which is designed to enhance information flow and feature fusion across different scales. PAFPN introduces a bottom-up path augmentation to the traditional FPN, enhancing low-level feature information and enabling better feature fusion. This allows for the combination of features from different stages of the backbone, thereby improving detection performance.
    (c) Head: 
    YOLOv8 introduces a decoupled head, which separates the classification branch from the localization branch. This decoupling alleviates the inherent conflict between classification and regression tasks, resulting in more accurate and reliable object detection. The decoupled head consists of a Detection Head and a Classification Head, which are responsible for predicting bounding boxes and class probabilities, respectively.
    }
    \label{fig:yolov8}
\end{figure}

YOLOv8 (You Only Look Once version 8), proposed by Ultralytics in 2023, is a state-of-the-art single-stage object detection model that enables end-to-end detection. Since the introduction of YOLO in 2016, YOLOv8 has been designed to provide fast and accurate object detection with several structural, efficiency, and performance improvements. As an end-to-end model, YOLOv8 simplifies the detection pipeline, reducing computational overhead and enabling automatic feature learning, which leads to better overall performance. This streamlined approach ensures that all components are optimized together, resulting in highly efficient training and inference, making YOLOv8 particularly well-suited for real-time applications.
The network structure is divided into three parts: the backbone, the neck, and the head, as shown in \autoref{fig:yolov8}.

\textbf{Characteristics of YOLOv8.}
\begin{itemize}
    \item Anchor-Free Mechanism: 
    YOLOv8 adopts an anchor-free mechanism that simplifies the model by reducing the number of hyperparameters and directly predicting the center coordinates of bounding boxes. This approach avoids anchor boxes predefined, reducing complexity and making the prediction process more efficient and straightforward.
    \item Decoupled Head: 
    YOLOv8 features a decoupled head structure that separates object classification from localization tasks. By addressing these tasks independently, the model optimizes each more effectively, allowing the network to specialize in accurately identifying object categories and precisely locating them within the image.
    \item Multi-Scale Predictions: 
    YOLOv8 is designed to make predictions at multiple scales, enhancing its ability to detect objects of various sizes. This multi-scale capability is achieved by making predictions at different stages of the network, each corresponding to varying levels of feature granularity. The Path Aggregation Network (PANet) in the neck plays a crucial role in enhancing multi-scale feature representation, allowing for better feature fusion and improving object detection across different scales.

    \item Advanced Feature Extraction with CSPDarknet: 
    YOLOv8 utilizes CSPDarknet as its backbone network, an advanced variant of the Darknet architecture with Cross Stage Partial (CSP) connections. These connections divide the feature map into two parts, merging them through a cross-stage hierarchy. This design improves gradient flow and reduces computational bottlenecks, leading to more efficient feature extraction. The CSPDarknet backbone keeps a balance between performance and computational efficiency, providing strong feature representations while maintaining high inference speed.  
\end{itemize}

\textbf{Training detail.}
The hardware setup includes an 11th Gen Intel(R) Core(TM) i7-11700 @ 2.50GHz and an NVIDIA GeForce RTX 3070.
The software environment consists of PyTorch v2.0.0 running on Windows 10. For end-to-end network optimization, we use the SGD algorithm with the official YOLOv8n pretrained weights, and the following hyperparameters: 

The learning rate is set to 0.01, weight decay to 0.0005, momentum to 0.937, and the batch size to 48. Training runs for 300 epochs, with early stopping triggered if performance does not improve after 100 epochs. All other configurations remain as the default settings of the original YOLOv8 model.

% Camera Parameter Acquisition, Calibration, and Time Synchronization
\subsection{
Camera Parameter Acquisition
\label{sec:preparation-steps-for-reconstruction}
}
The camera parameters, including resolution, frame rate, intrinsic parameters, distortion coefficients, position, corresponding frames, and unified time, depend on the specific camera model (as detailed in \autoref{tab:camera-specifications}).

The resolution and frame rate are obtained from the camera technical specifications (we use an iPad as the camera, with 1080p resolution (1920 x 1080 pixels) and a frame rate of 30 fps; its specifications can be found on the official website at \url{https://www.apple.com/ipad/specs/}).

\begin{figure}[htbp]
    \centering
    \includegraphics[width=\linewidth]{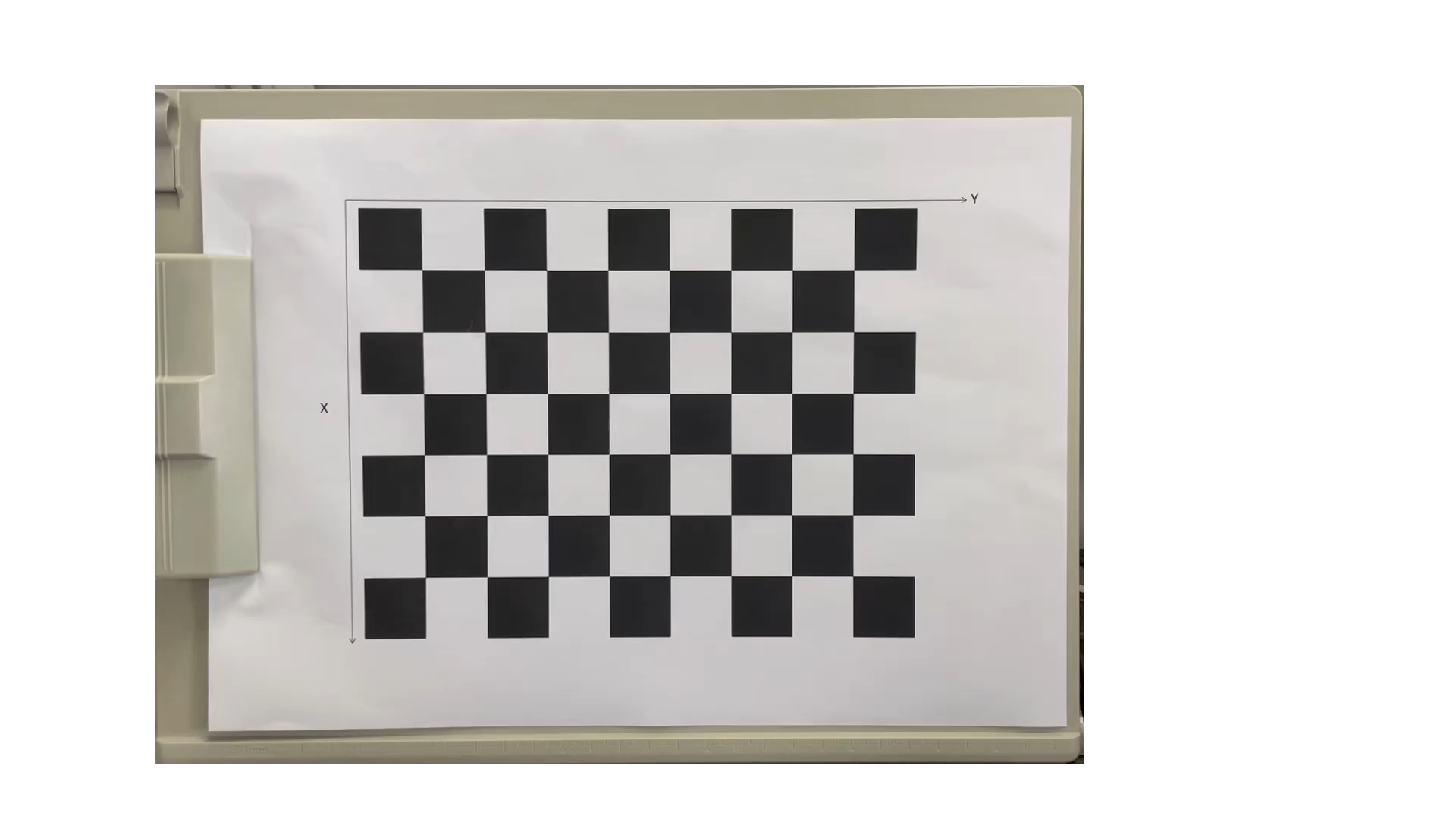}
    \caption{
    Calibration plate used for intrinsic parameter calibration, featuring a 7x13 grid with each square measuring 21 mm.}
    \label{fig:calibration-plate}
\end{figure}

The intrinsic parameters and distortion coefficients are estimated using the Zhang Zhengyou method \cite{zhang2000flexible}, a widely recognized approach for camera calibration. This method involves capturing a series of images from different angles to accurately calculate these parameters. The calibration tool used is a calibration plate, as shown in \autoref{fig:calibration-plate}, where each square on the plate measures 21 mm. During calibration, the plate is positioned at various orientations and distances from the camera, ensuring a comprehensive calibration across different perspectives.

Camera positions are measured using a high-precision RTK GNSS receiver, providing each camera position within the world coordinate system with an accuracy within 10 cm. 
We use WGS84 as our geodetic reference ellipsoid, which provides components along the X, Y, and Z axes in the Cartesian coordinate system.
However, an installation error also exists between the GNSS receiver and the camera optical center, estimated at approximately 10 cm. 
These high-precision RTK GNSS devices are connected to a local Continuously Operating Reference Stations (CORS) network, which provides GNSS positioning services across the area. 
The CORS network enables RTK GNSS systems to achieve 3D positioning accuracy of approximately $\pm 3\,\text{cm}$ under optimal conditions \cite{alkan2020comparative}.

\begin{figure}[htbp]
    \centering
    \includegraphics[width=\linewidth]{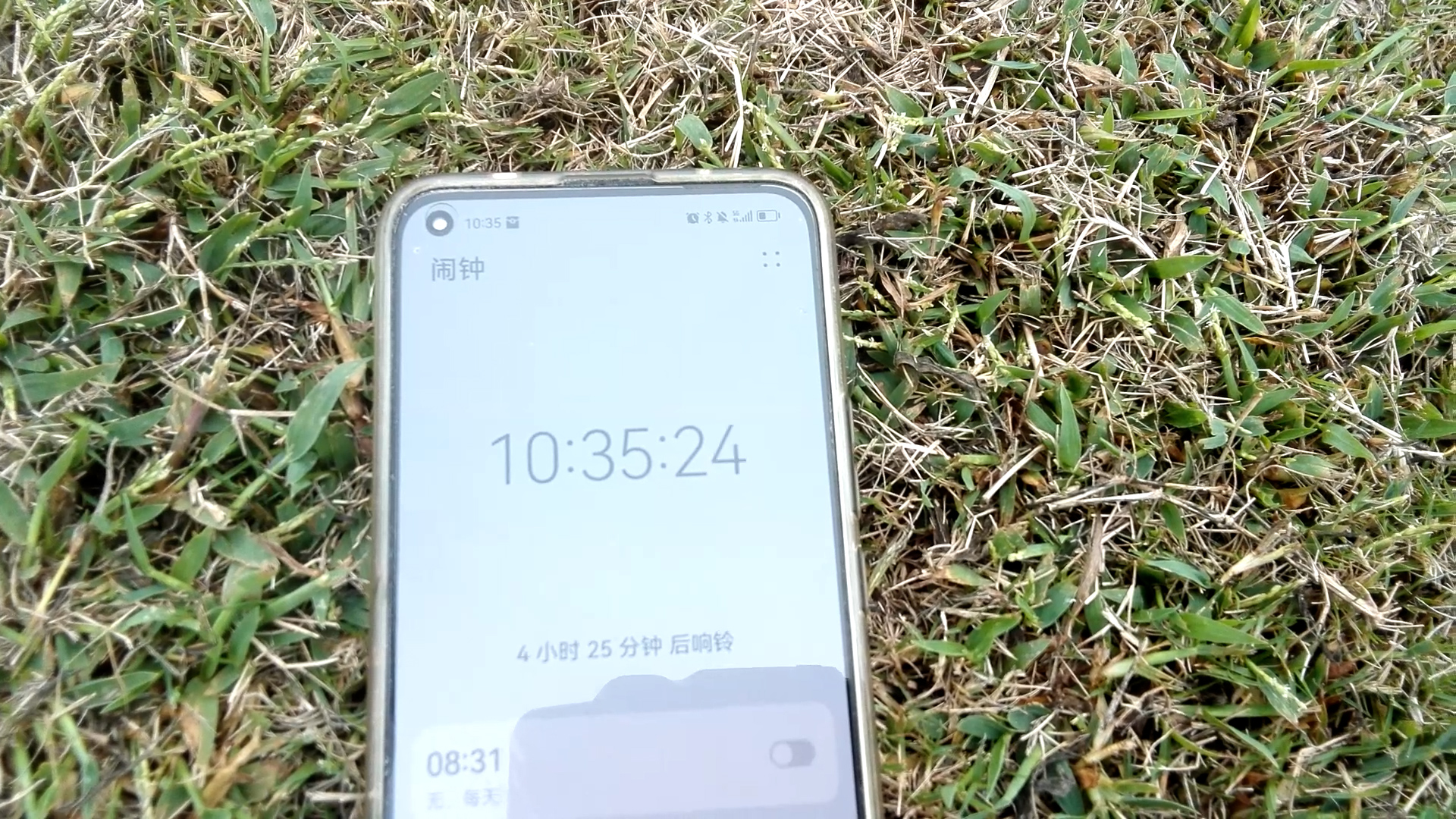}
    \caption{
    Relationship between camera frames and unified time. This camera records frame 34172 at 10:35:24 Beijing time.
    % (a) Camera 1 records frame 34172 at 10:35:24 Beijing time. (b) Camera 3 records frame 46683 at 10:40:49 Beijing time. (c) Camera 5 records frame 44893 at 10:38:29 Beijing time.
    }
    \label{fig:frame_beijingtime}
\end{figure}

We establish the relationship between video frames and the time reference (Beijing time in this case) for camera time synchronization by having each camera continue recording while capturing the Beijing time. As shown in \autoref{fig:frame_beijingtime}, camera1 records frame 34172 at 10:35:24 Beijing time. Similarly, camera3 records frame 46683 at 10:40:49, and camera5 records frame 44893 at 10:38:29.

% \begin{table*}[!ht]
\begin{sidewaystable}
\centering
\caption{Parameters for multiple camera models}
\small
\begin{tabular}{ccccccccc}
\toprule
\makecell{camera\\index} & Model & Resolution & \makecell{Frame\\rate}  & Intrinsic Parameters Matrix & \makecell{Distortion\\Coefficients} & \makecell{Position \\(m)} & \makecell{Corresponding\\Frames} &  unified time \\
\midrule
0 & HUAWEI & 1920x1080 & 30  & \(\begin{bmatrix}
1599.9 & 0 & 962.3 \\
0 & 1601.3 & 560.7 \\
0 & 0 & 1
\end{bmatrix}\) & \(\begin{bmatrix}
0.1084 \\ 0.0636 \\  0.0040 \\  0.0048 \\ -0.5204
\end{bmatrix}\)  
& (37.6, 127.4, 0.1) & 34172 & 10:35:24 \\
\midrule
1 & HUAWEI & 1920x1080 & 29.99  & \(\begin{bmatrix}
1603.1 & 0 & 960.0 \\
0 & 1603.9 & 566.0 \\
0 & 0 & 1
\end{bmatrix}\) & \(\begin{bmatrix}
0.1169 \\ -0.2359 \\  0.0050 \\  0.0035 \\ 1.0072
\end{bmatrix}\) 
& (81.4, 24.4, 0.1) & 46683 & 10:40:49 \\
\midrule
2 & HUAWEI & 1920x1080 & 30  
& \(\begin{bmatrix}
1655.9 & 0 & 977.1 \\
0 & 1657.5 & 549.7 \\
0 & 0 & 1
\end{bmatrix}\) &  \(\begin{bmatrix}
0.0735 \\ 0.4812 \\  0.0034 \\  0.0043 \\ -2.4223
\end{bmatrix}\) 
& (15.3, 19.6, 0.1) & 44893 & 10:38:29 \\
\bottomrule
\end{tabular}
\label{tab:camera-specifications}
\end{sidewaystable}
% \end{table*}

\subsection{Preparation for UAV coordinate recognition
\label{sec:preparation-for-UAV-coordinate-recognition}
}
After obtaining the detection boxes for all cameras at each time step, we calculate the center coordinates of these detection boxes as the UAV 2D coordinates in the pixel coordinate systems of the three cameras.

\textbf{Preliminary preparation.}
The 2D pixel coordinates of the UAV captured by each camera at each time step cannot be directly used for 3D coordinate recognition due to several factors: 
geometric distortion from the lens, undetermined 3D coordinates of the cameras in the world coordinate system, unknown intrinsic camera parameters, and lack of time synchronization between the image series from different cameras.
To address these issues, preliminary steps include calibrating the intrinsic parameters and correcting for distortion using Zhang Zhengyou method \cite{zhang2000flexible}, determining the camera coordinates in the world coordinate system with a GNSS receiver, and synchronizing the image time series from different cameras (see Supplementary Information Section~\ref{sec:preparation-steps-for-reconstruction}).

\textbf{Calibration and Synchronization.}
Using the distortion parameters of the camera (obtained from Supplementary Information Section~\ref{sec:preparation-steps-for-reconstruction}), we correct the deviations in the 2D pixel coordinates of the UAV. 
Following this, the time correspondence between different cameras is established, allowing us to convert the 2D pixel coordinate time series from the original camera-based time to a unified global time.

\textbf{Trajectory continuation and interpolation.}
In practice, the 2D coordinate time series of the UAV obtained from camera 0 and camera 1 are unaligned due to the discrete nature of image capture.
Additionally, the UAV 2D coordinates may not be captured at every time step if the UAV flies out of the camera field of view or if the YOLOv8-TS model fails to detect the UAV. Thus we first identify the continuous sub-time series during overlapping periods when both cameras successfully capture the UAV.
To align these time series, we smooth the UAV 2D coordinate time series from camera 1 using cubic spline interpolation \cite{schoenberg1988contributions} and resample it at the time points of time series from camera 0, generating an aligned timeline.
For the remaining cameras \(i\) (where \(N > i \geq 2\); \(i = 2\) and \(N = 3\) in the UAV experiment), we identify the overlapping time periods between the initial 3D trajectory and the time when camera \(i\) observes the UAV. We then interpolate the 2D coordinate time series from camera \(i\) to match the time points of the initial 3D trajectory.
Additionally, because the initial UAV 3D coordinate time series is often shorter than the total flight time, the 2D coordinate time series from the remaining cameras can be used to extend the UAV 3D coordinates for the remaining periods (\autoref{fig:UAV-experiment-schematic-diagram}).

\section*{Acknowledgment}

The author, Ke-ke Shang, dedicates this work as a tribute to the traditional research direction of his doctoral major.  While he has since embarked on an interdisciplinary path driven by a deeper passion for the social sciences, he remains profoundly grateful for the rigorous academic foundation and intellectual training provided by his doctoral program.

This paper stands not only as a continuation of scholarly inquiry but also as a mark of respect and gratitude toward the field that shaped his early academic identity. To the mentors, colleagues, and institutions that supported him during those formative years: thank you.

May this work honor both the past that guided him and the future he continues to explore.

\section*{Contributions}
\begin{itemize}
    \item \textbf{Junfan Yi} (Co-first author): Experiment design, Linear algebra analysis, Code implementation, Data analysis, Experiment execution, Software operation, Initial draft writing.
    \item \textbf{Ke-ke Shang} (Co-first author \& Corresponding author): Linear algebra analysis, Experiment design, Data analysis, Simulation design, Supervision, Writing, Reviewing.
    \item \textbf{Michael Small}: Writing, Supervision, Reviewing.
\end{itemize}

  \bibliography{ref}
  \bibliographystyle{ieeetr}
\end{document}